%% file: main_codaspy.tex
\documentclass[sigconf]{acmart}

\AtBeginDocument{%
  \providecommand\BibTeX{{%
    \normalfont B\kern-0.5em{\scshape i\kern-0.25em b}\kern-0.8em\TeX}}}

\usepackage{tikz}
\usepackage{amsmath}
\usepackage{booktabs}
\usepackage{hyperref}
\usepackage{amsfonts}
\usepackage{algorithmic}
\usepackage{xspace}
\usepackage{enumitem}
\usepackage{graphicx}
\usepackage{textcomp}
\usepackage{color}
\usepackage{xcolor}
\usepackage{subfigure}
\usepackage{caption}
\usepackage[normalem]{ulem}
\usepackage{comment}
\usepackage{filecontents}
\usepackage{multirow}
\usepackage{makecell}
\usepackage{url}
\newcommand{\says}[2]{{\color{blue}{#1 says: }{#2}}\xspace}
\newcommand{\mypara}[1]{\vspace*{0.06in}\noindent\textbf{#1 }}
\newcommand{\norm}[1]{\left\lVert#1\right\rVert}

\setcopyright{acmcopyright}

\copyrightyear{2021} 
\acmYear{2021} 
\setcopyright{acmcopyright}\acmConference[CODASPY '21]{Proceedings of the Eleventh ACM Conference on Data and Application Security and Privacy}{April 26--28, 2021}{Virtual Event, USA}
\acmBooktitle{Proceedings of the Eleventh ACM Conference on Data and Application Security and Privacy (CODASPY '21), April 26--28, 2021, Virtual Event, USA}
\acmPrice{15.00}
\acmDOI{10.1145/3422337.3447836}
\acmISBN{978-1-4503-8143-7/21/03}

\usepackage{filecontents}

\begin{document}
\fancyhead{}
%
\title{Membership Inference Attacks and Defenses in Classification Models}

\author{Jiacheng Li}
\affiliation{%
  \institution{Purdue University}
  \city{West Lafayette}
  \state{Indiana}
  \country{USA}
}
\email{li2829@purdue.edu}
\author{Ninghui Li}
\affiliation{%
  \institution{Purdue University}
  \city{West Lafayette}
  \state{Indiana}
  \country{USA}
}
\email{ninghui@cs.purdue.edu}
\author{Bruno Ribeiro}
\affiliation{%
  \institution{Purdue University}
  \city{West Lafayette}
  \state{Indiana}
  \country{USA}
}
\email{ribeiro@cs.purdue.edu}

\begin{abstract}
We study the membership inference (MI) attack against classifiers, where the attacker's goal is to determine whether a data instance was used for training the classifier.  Through systematic cataloging of existing MI attacks and extensive experimental evaluations of them, we find that a model's vulnerability to MI attacks is tightly related to the generalization gap---the difference between training accuracy and test accuracy.  
We then propose a defense against MI attacks that aims to close the gap by intentionally reduces the training accuracy.  More specifically, the training process attempts to match the training and validation accuracies, by means of a new {\em set regularizer} using the Maximum Mean Discrepancy between the softmax output empirical distributions of the training and validation sets.  
Our experimental results show that combining this approach with another simple defense (mix-up training) significantly improves state-of-the-art defense against MI attacks, with minimal impact on testing accuracy. 
\end{abstract}



\maketitle

\input{1_intro.tex}

\input{2_background.tex}

\input{3a_existing_attacks.tex}

\input{3b_existing_eval.tex}

\input{4a_defenses.tex}

\input{4b_defense_eval.tex}

\input{6_related_work.tex}

\input{7_conclusion.tex}

\bibliographystyle{ACM-Reference-Format}
\bibliography{privacy}

\section{Acknowledgments}
This work is supported in part by the United States National Science Foundation under Grant No. 1931443.
\input{8_appendix.tex}

\end{document}

%% file: 1_intro.tex
\section{Introduction}
\label{sec:intro}
    

    In recent years we see widespread adoption of Machine Learning (ML) techniques in real-world applications, such as image and speech recognition, and user behavior prediction. Increasingly, as companies and organizations integrate ML components into their products and services, this ubiquitous deployment also leads to increasing scrutiny on the security and privacy implications of ML techniques. 

    The threat of membership inference (MI) attacks against classifiers --- where an attacker detects whether a specific data instance was used as training data--- has received much attention in the literature ~\cite{shokri2017membership,long2017towards,yeom2018privacy,nasr2018machine,nasr2018comprehensive}. 
    An attacker---who is given access to the target classifier--- seeks to determine whether an instance $(x,y)$---where $x$ is the feature vector and $y$ is its label--- \textbf{is a member} of the dataset used to train it (the classifier).   
    An MI attack is itself a binary classification, and will have false positive and false negative rates depending on the set of instances used in the evaluation.  A standard way used in the literature~\cite{shokri2017membership,long2017towards,yeom2018privacy,nasr2018comprehensive} to measure the effectiveness of an MI attack and compare different attacks, is to measure its accuracy over a balanced set that consists of half members (instances used in training) and half non-members.  Since a trivial attack of randomly guessing whether an instance is a member has accuracy of $1/2$, we define the {\em advantage} of an attack to be its accuracy over a balanced set minus $1/2$.

    Effective MI attacks pose a significant threat to the privacy of the individuals whose records have been used in training data. For example, cancer patient records may be used to train a classifier to predict what kind of treatment would be most promising for a future patient. However, if high-accuracy MI attacks can be carried out against this classifier, access to the classifier may enable adversaries to learn sensitive personal information about patients whose data were used as training data.  

    
    It is known that overfitted classifiers are more susceptible to MI attacks~\cite{shokri2017membership,salem2018ml,yeom2018privacy}.  Given a classifier, its \textbf{generalization gap} $g$ is defined to be $g=a_R-a_E$, where $a_R$ is the classifier's accuracy on training data, and $a_E$ is the classifier's accuracy on a dataset drawn from the same distribution as the training data, e.g., the testing data.  A classifier that overfits has a large generalization gap. 
    Intuitively, the generalization gap of a classifier is closely related to its MI attack vulnerability.  A baseline attack ---first introduced by Yeom et al.~\cite{yeom2018privacy}---concludes that an input $(x,y)$ is a member of the training data if and only if the classifier yields the correct label $y$ on $x$.  
    The accuracy of the baseline attack for a balanced evaluation set is $\frac{1}{2}*a_R + \frac{1}{2}*(1-a_E) = \frac{1+a_R-a_E}{2} = \frac{1+g}{2}$.  Its advantage is therefore $\frac{g}{2}$.  Thus, as long as the generalization gap exists, the baseline attack will have some level of effectiveness. 

\mypara{Contributions.}    In this paper, we study MI attacks and defenses, and make the following contributions:
\begin{enumerate}

\item We systematically catalog existing MI attacks and perform extensive experimental evaluations of their effectiveness.  From the experiments, we find that the simple baseline attack provides an excellent estimate of a model's vulnerability to existing MI attacks.  For many settings, the baseline attack achieves an advantage that is close to the highest advantage achieved by any existing MI attack.  In all settings (except for the MNIST dataset), the highest attack advantage achieved in any attack is less than 1.5 times that of the baseline attack. Models for the MNIST dataset are minimally affected by MI attacks, with the highest attack advantage being 2\%, and the baseline advantage being 1\%.

These results directly link the generalization gap with the vulnerability against MI attacks.  Any model is vulnerable to the baseline attack, which has an advantage of $g/2$.  And no MI attack performs significantly better than the baseline attack.  Understanding this link will help research on MI attacks.  
For example, some researchers have proposed to defend against MI attacks by limiting access to the prediction probability vector, e.g., restricting the probability vector to only the top $k$ classes, coarsen the precision of the probability vector, or increase the entropy by using temperature in softmax function.  These defenses are ineffective against the baseline attack. 
Any proposed defenses that do not close the generalization gap cannot be effective, even if they are able to reduce the advantage of certain attacks, as they remain vulnerable to the baseline attack. 

\item Our main contribution is a new, principled defense against MI attacks.  
%
We propose to \textbf{intentionally reduce training accuracy} to validation accuracy, while maintaining test accuracy.  We achieve this by adding a regularizer to the training loss function. Our new {\em regularizer} computes the distance between  the empirical distribution of the softmax output over the training set and that of a validation dataset.  It penalizes the situation where training instances have significantly different output distributions from the validation instances that are not directly used in training.  
Our penalty uses the \textbf{Maximum Mean Discrepancy (MMD)}~\cite{fortet1953convergence}, in the form of a non-parametric kernel two-sample test as introduced by Greton et al.~\cite{gretton2012kernel}, which gives us a measure of the difference between the two empirical distributions.  
In practice, MMD alone tends to reduce both the training and test accuracy, an unwanted outcome.  
To tackle this challenge, we propose to combine MMD with \textbf{mix-up  training}~\cite{zhang2018mixup}, which, in our experiments, garners the benefits of MMD without the test accuracy penalty.

 \item 

We empirically compare our proposed MMD+Mix-up regularizer defense with other defenses such as Mem-Guard~\cite{jia2019memguard} and DP-SGD.  For DP-SGD, we vary the amount of added noises and measure the resulting accuracy numbers as well as the highest attack advantage numbers.  We show that our proposed MMD+Mix-up defense achieves significantly better accuracy --- MI vulnerability tradeoff than existing defenses.

\end{enumerate}

\mypara{Organization.} The rest of this paper is organized as follows. Section~\ref{sec:adv_model} discusses adversary models and nomenclature of attacks. Section~\ref{sec:attack} summarizes and evaluates all existing attacks. Section~\ref{sec:defense} describes existing defenses and our proposed defenses. Section~\ref{sec:eval} presents the evaluation of our defense and existing defenses.
Related work is discussed in Section \ref{sec:related}. Section \ref{sec:conclusion} concludes this paper. An appendix gives additional information.

%% file: 2_background.tex
\section{Adversary Models and Nomenclature}\label{sec:adv_model}

\begin{table*}
    \small
    \centering
    \begin{tabular}{lrrrrr}
    Attacks  & Feature & Method  & Adversary Model & Computational Cost  \\
    \toprule
    
    Instance-Vector ~\cite{long2017towards} & Probability Vector $F^T(x)$ & KL distance to avg & Training plus Data & High \\
    
    Class-Vector ~\cite{shokri2017membership} & Probability Vector $F^T(x)$ & Neural Network & Training plus Data & Medium \\
    
    Global-Probability ~\cite{yeom2018privacy} & Probability of correct label & Threshold & Training plus Data & Medium\\
    
    Global-Loss ~\cite{yeom2018privacy} & Training Loss $Loss(x,y)$ & Threshold & Training plus Data & Medium\\
    
    Global-TopThree ~\cite{salem2018ml} & Top3 in $F^T(x)$ & Neural Network & Training plus Data & Medium \\

    Global-TopOne ~\cite{salem2018ml}& Top1 in  $F^T(x)$ & Threshold & Probability-Vector Oracle & Low\\
    
    Baseline ~\cite{yeom2018privacy} & Predicted Class $y'$ & Binary & Label-only Oracle & Very Low\\
    
    \bottomrule
    
    \end{tabular}
    \caption{Summary of different black-box MI attacks proposed in previous works.}
    \label{tab:attack_model_summary}
\end{table*}

We consider ML models that are used as $c$-class classifiers.  Each instance is a pair $(x,y)$, where $x$ is the feature and $y$ is the label.  A classifier typically uses a softmax function and is trained with cross-entropy loss function.  
For any input instance $x$, the target classifier $F^T$ outputs $F^T(x)$, a length-$c$ vector of non-negative real values that are outputs of the softmax function.  The components of $F^T(x)$ sum up to 1 and are generally interpreted as the confidence values that $x$ belongs to each of the $c$ classes.  The predicted class is the one with the highest corresponding value in $F^T(x)$.  
In some cases, the values in the vector $F^T(x)$ are calibrated so that they indeed represent some probabilities; in other cases, these values sum up to $1$, but do not have real probability interpretation.  For convenience, we call $F^T(x)$ the \emph{probability vector}.  

The adversary's goal is to determine whether an instance $x$ is a member of the dataset for training a target classifier $F^T$.  We consider the following three classes of adversaries, in increasing strengths.  

\begin{itemize}
    \item \textbf{Label-only Oracle.}
    The adversary has no knowledge (or, equivalently, does not utilize any such knowledge) of the training process or the data distribution, and is given only the predicted label of each instance.  
    
    Some proposed defenses against MI attacks include limiting the amount of information in the predicted probability vector.  This model is under the extreme situation where the probability vector is completely removed.  
    The only attack that is feasible under this model is the baseline attack, which predicts one instance to be a member when the target model returns the correct label.


    \item \textbf{Probability-Vector Oracle.}
    The adversary is given blackbox/oracle access to the target classifier, which gives a probability vector for each instance.  
    This is perhaps the most natural adversary model. 

    \item \textbf{Training-plus-Data.}
    The adversary knows the training process such as the classifier architecture, hyperparameters, and training algorithms, and the distribution of the data used in training.  In addition, the adversary has oracle access to the target classifier $F^T$, that is, the adversary can query $F^T$ and obtain the probability vector.  

    Knowledge of the training process and data distribution enables the adversary to train \emph{shadow classifiers} that behave similarly to the target classifier and learn information about the target classifier through the shadow classifiers.  This adversary may not be realistic in all settings; however, understanding a model's vulnerability to MI attacks under this model provides an upper-bound. 
\end{itemize}

    In~\cite{nasr2018comprehensive}, other adversary models for MI attacks were considered.  These include white-box attacks, in which the attacker also knows the parameters of the target classifier, and federated training settings, in which the attacker may also observe the updates during the training.  
    Similar to~\cite{salem2018ml,jia2019memguard,carlini2018text,long2017towards,long2018understanding,nasr2018machine}, we focus on the setting where the adversary uses oracle accesses to the target classifier, although a Training-plus-Data attacker has access to sufficient information to train shadow classifiers similar to the target classifier.   
    Threats from attacks in the black-box settings are more serious because it can be carried out by anyone who can query a classifier, e.g., in the Machine Learning as a Service setting.  Understanding threats from MI attacks in the black-box setting also helps understanding MI attacks in other settings.

%% file: 3a_existing_attacks.tex
\section{Evaluating Existing MI Attacks} \label{sec:attack}

In this section, we discuss existing membership inference (MI) attacks, empirically evaluate their effectiveness, and compare these with the generalization gap.   These results help us design defenses against MI attacks in Section~\ref{sec:defense}.


\subsection{Summary of MI Attacks}

We now summarize black-box MI attacks that have been proposed in the literature.  We assign a two-part name to each attack.  
The first part of a name is based on the \emph{granularity} of the MI attack.  An attack may use one \textbf{Global} model or threshold value for all instances, or use one model (or threshold value) for each \textbf{Class} or each \textbf{Instance}. 
The second part of a name is based on the \emph{features} used in the attack.  

\mypara{The Class-Vector Attack~\cite{shokri2017membership}.}
    To our knowledge, Shokri et al.~\cite{shokri2017membership} presented the first study on MI attacks against classifiers. The attack is in the ``Training-plus-data'' adversary model, and uses the probability vector $F^T(x)$ as the feature for determining whether $x$ is a member.  
    
    The adversary knows a dataset $D^A$, which is from the same distribution as the dataset used to train the target classifier. 
    The adversary creates $k$ samples $D_1,D_2,\ldots,D_k$ from $D^A$, and trains $k$ shadow classifiers, one from each $D_i$.
    Using data generated by these shadow classifiers, the attacker trains $c$ MI classifiers, one for each class. The classifier for class $y$ is trained using instances in $D^A$ that are of class $y$.   
    Each MI classifier takes a probability vector $F^T(x)$ as input, and produces a binary classification result.  
    When trying to determine the membership of an instance $(x)_y$, one feeds $F^T(x)$ to the MI classifier for class $y$ to obtain a binary membership prediction. The computational cost of this method includes training of multiple shadow models and training of attack models.

\mypara{The Baseline (Global-Label) Attack ~\cite{yeom2018privacy}.}  
Yeom et al.\cite{yeom2018privacy} analyzed the relationship between overfitting and membership, and proposed two attacks, one of which predicts that an instance $x$ is a member for training $F^T$ if and only if $F^T(x)$ gives the correct label on $x$. This attack can be applied to the ``Label-only Oracle'' adversary model, in which the adversary is given only the label. 


The advantage of the Baseline attack can be estimated from the training and testing accuracy.  
Let $a_R$ be the training accuracy, and $a_E$ be the testing accuracy.  Then $g=a_R-a_E$ is the \textbf{generalization gap}.   
Given a balanced evaluation set, the accuracy of the baseline attack is the average of its accuracy on members and non-members.  By definition, its accuracy on members is about $a_R$ and its accuracy on non-members is about $1-a_E$.  Its overall accuracy is thus $\frac{1}{2}*a_R + \frac{1}{2}*(1-a_E) = \frac{1+a_R-a_E}{2} = \frac{1+g}{2}$. Its advantage is therefore $\frac{g}{2}$.   This estimation of accuracy is an approximation but our experiments empirically show that this estimation is quite accurate.
Since this attack is so simple and broadly applicable, we call this attack the baseline attack.  This attack's advantage provides a lower-bound estimation of a classifier's vulnerability to MI attacks.  The computational cost of this attack is minimal, since to determine the membership of an instance, this attack just needs one query on the target classifier.


\mypara{The Global-Loss Attack \cite{yeom2018privacy}.}
This attack uses the probability vector $F^T(x)$ for an instance $x$ with true label $y$ to compute the cross-entropy loss: 
        $Loss(x,y) = -\log(F^{T}(x)_y)$,
    where $F^{T}(x)_y$ is the probability value for the true label $y$. The attack predicts $x$ to be a member when $Loss(x,y)$ is smaller than the average loss of all training instances.  
    We consider this attack to be in the Training-plus-Data adversary model, because the average training loss is not normally provided. The natural way to obtain it is to build one or more shadow classifiers. This method also requires training one or more shadow models, from which one can estimate the average loss of all training instances in the target model.
    
\mypara{The Global-Probability Attack.}
We note that the Global-Loss attack effectively predicts an instance to a member if the probability for the correct label is above some threshold.  It fixes the threshold based on the average value for all training instances.      This threshold may not achieve the maximum accuracy.  We thus also consider using shadow classifiers and training data to compute the threshold that achieves the best accuracy.  We call this the Global-probability attack. This method needs to train shadow models and the best threshold needs to be calculated.
    
    
    
\mypara{The Global-TopOne Attack \cite{salem2018ml}.}
Instead of using the probability of the correct label, Salem et al.~\cite{salem2018ml} proposed to use the highest value in the probability vector.  
They proposed an interesting threshold-choosing approach that exploits oracle access to the target classifier.  One randomly generates some data instances, which are non-members with high probability, and query the target classifier with these instances.  One then chooses the threshold using the top $t$ percentile among the Top 1 probabilities from the probability vectors of these instances. Experiments on different $t$ in a range from $5\%$ to $25\%$ showed a decent performance. The computational cost of this method is moderate since model training is not needed and the adversary only needs to make some queries to select the threshold.

\mypara{The Global-TopThree Attack \cite{salem2018ml}.}
    Salem et al.~\cite{salem2018ml} proposed a data transferring attack, where the adversary trains one shadow classifier using a different dataset and different classifier structure. Since the number of classes may be different for the shadow classifier and the target classifier, the adversary chooses the top 3 values in the probability vector (top 2 in case of binary classification) as the features for MI attack.  Furthermore, only a single global MI classifier is used. The computational cost of this method includes training of shadow models using different datasets and training of attack models. The attack model can be logistic regression or neural network.

\mypara{The Instance-Vector Attack~\cite{long2017towards}.}
    Long et al.\cite{long2017towards} proposed three MI attacks, all in the ``Training-plus-data'' adversary classifier.  
    The first attack in \cite{long2017towards} is called the untargeted attack, which trains an MI classifier that takes a probability vector $F^T(x)$ and the label $y$ as the feature, and whether $(x,y)$ is used in training the classifier $M$ as the label.  This is essentially equivalent to the Class-Vector attack discussed above.   Even though this attack trains just one classifier, the class label is taken as an input into the classifier, enabling the classifier to adapt based on the class label.  Because it is essentially equivalent to the Class-Vector attack, we do not consider it separately in the rest of this paper. 
    
    
    
    The remaining two attacks in \cite{long2017towards} train instance-specific MI classifiers. That is, there is one MI classifier for each instance $x$.  To enable this, one creates $k$ samples $D_1,D_2,\cdots,D_k$ of $D^A$, and trains $2k$ shadow classifiers, where $F^S_i$ is trained with $D_i$, and $F'^{S}_i$ is trained with $D_i \cup \{x\}$.  For each instance $x$, one has $k$ probability vectors  $F^S_i(x)$, where $1\le i\le k$, from classifiers trained without $x$, and $k$ probability vectors  $F'^S_i(x)$, from classifiers trained with $x$.  The intuition is that if $x$ is used in training $F^T$, then the probability vector $F^T(x)$ should be more similar to the latter $k$ than to the former $k$.  
    
    Long et al.~\cite{long2017towards} investigated two ways to measure this similarity, and found that the more effective approach is to use the Kullback-Leibler (KL) divergence.  
    This attack requires training of shadow models and the number of shadow models needed is significantly more since this attack is an instance-based attack. For each training instance, the number of training points that are available to the attack model is the same as the number of shadow models. Thus, a large number of shadow models are preferred.

%% file: 3b_existing_eval.tex
\vspace{-0.6em}
\subsection{Experimental Setup} \label{sec:experimental_setup_1}

We experimentally evaluate the effectiveness of these existing attacks.  Here we first describe the experimental setup. 

\begin{figure*}
     \centering
     \includegraphics[width=18cm,height=9cm]{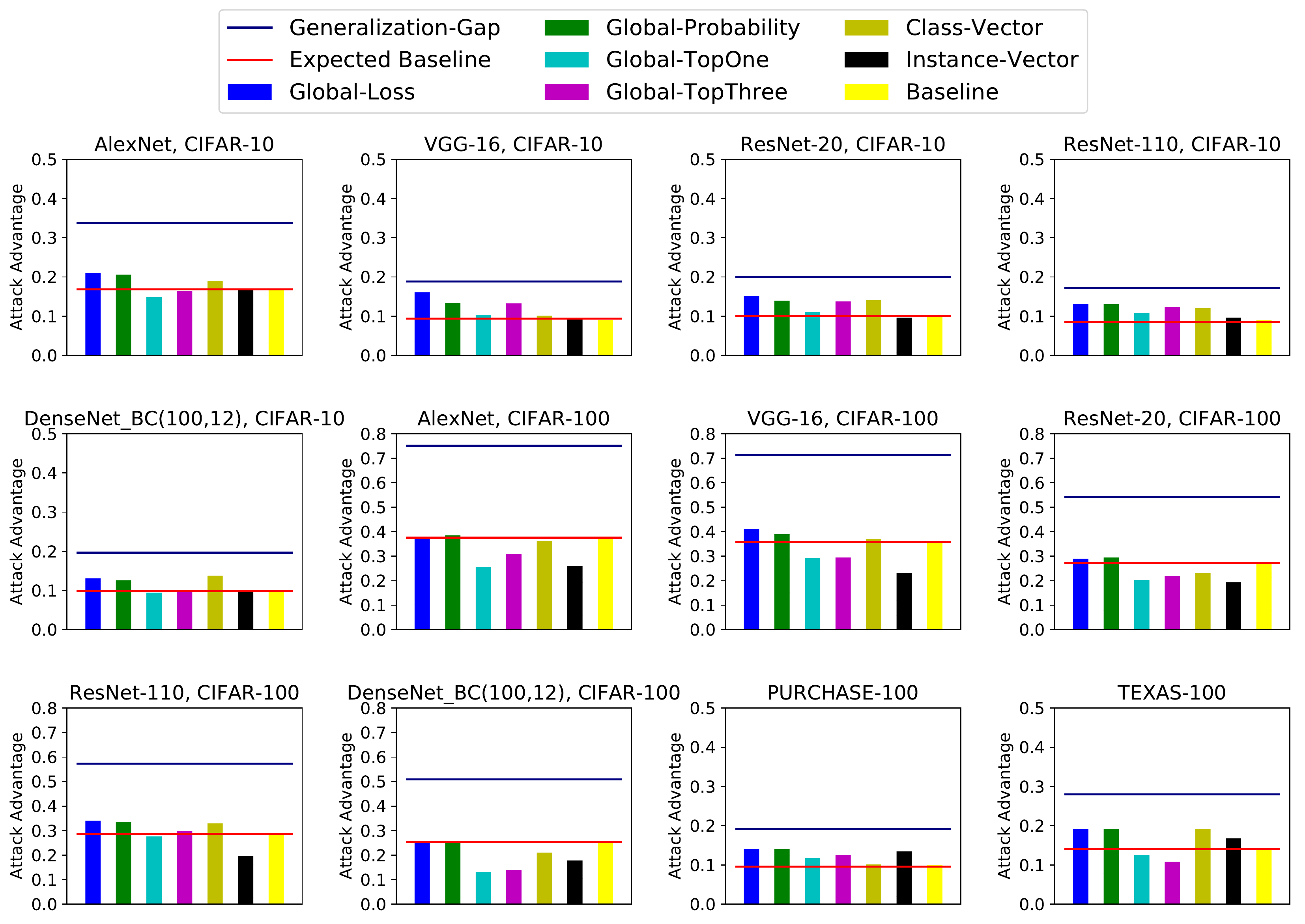}
     \caption{Attack advantage of different attacks for 5 models each on CIFAR-10 and CIFAR-100, and 1 model each for PURCHASE-100 and TEXAS-100. Each bar represents a different attack. The higher horizontal line represents the generalization gap $g$ and the lower horizontal line represents the expected baseline attack advantage ($g/2$).}
     \label{disjoint}
 \end{figure*}
 
\begin{table*}
    \normalsize
    \centering
    \begin{tabular}{lrrrrr}
        dataset & CIFAR-10 & CIFAR-100 & TEXAS-100 & PURCHASE-100 & MNIST \\
        \toprule
        Training accuracy & 0.994 & 0.984 & 0.815 & 0.981 & 0.999 \\
        Testing accuracy & 0.776 & 0.366 & 0.535 & 0.791 & 0.981 \\ 
        \cmidrule(lr){1-6}
        \textbf{Generalization gap} & {\textbf{0.218}} & {\textbf{0.618}} & {\textbf{0.280}} & \textbf{0.190} & \textbf{0.018} \\
        \textbf{Largest attack advantage} & \textbf{0.156} & \textbf{0.332} & \textbf{0.191} & \textbf{0.145} & \textbf{0.020}  \\
        \textbf{Baseline attack advantage} & \textbf{0.112} & \textbf{0.308} & \textbf{0.143} & \textbf{0.101} & \textbf{0.010} \\
        \cmidrule(lr){1-6} 
        Class-Vector attack advantage & 0.138 & 0.299 & 0.161 & 0.106 & 0.013 \\
        Global-Loss attack advantage & 0.156 & 0.331 & 0.191 & 0.143 & 0.013 \\
        Global-Probability attack advantage & 0.147 & 0.332 & 0.191 & 0.145 & 0.013 \\
        Global-TopOne attack advantage & 0.113 & 0.231 & 0.125 & 0.116 & 0.011 \\
        Global-TopThree attack advantage & 0.131 & 0.251 & 0.108 & 0.125 & 0.012 \\
        Instance-Vector attack advantage & 0.111 & 0.211 & 0.167 & 0.133 & 0.020 \\
        \bottomrule
    \end{tabular}
    \caption{Training/testing accuracy and attack advantage for different datasets. For CIFAR-10 and CIFAR-100, we average throughout all the tested classifiers.}
    \label{disjoint_avg_attack_advantage}
\end{table*}

\mypara{Datasets.}  We use the following datasets, which are used in existing work on MI attacks. 

    \mypara{CIFAR-10 and CIFAR-100.}  These are benchmark datasets for evaluating image classification algorithms and systems. They contain 60,000 color images of size 32 $\times$ 32, divided into 50,000 for training and 10,000 for testing.  In CIFAR-100, these images are divided into 100 classes, with 600 images for each class.  In CIFAR-10, these 100 classes are grouped into 10 more coarse-grained classes; there are thus 6000 images for each class. These two datasets are widely used to evaluate membership inference attack in ~\cite{shokri2017membership,salem2018ml,nasr2018machine,nasr2018comprehensive,yeom2018privacy}.
    
    
    
    \mypara{PURCHASE-100.} This dataset is based on the ``acquire valued shopper'' challenge from Kaggle. This dataset includes shopping records for several thousand individuals.  We obtained the processed and simplified version of this dataset from the authors of~\cite{shokri2017membership}.  Each data instance has 600 binary features. This dataset is clustered into 100 classes and the task is to predict the class for each customer. The dataset contains 197,324 data instances. This dataset is also widely used to evaluate membership inference attack in  ~\cite{shokri2017membership,salem2018ml,nasr2018machine,nasr2018comprehensive,jia2019memguard}.
    
    \mypara{TEXAS-100.} This dataset includes hospital discharge data. The records in the dataset contain information about inpatient stays in several health care facilities published by the Texas Department of
    State Health Services. Data records have features about the external causes of injury, the diagnosis, the procedures the patient underwent, and generic information. We obtained a processed version of the dataset from the authors of \cite{shokri2017membership}.  This dataset contains 67,330 records and 6,170 binary features which represent the 100 most frequent medical procedures. The records are clustered into 100 classes, each representing a different type of patient. This dataset is used to evaluate membership inference attack in  ~\cite{shokri2017membership,nasr2018machine,nasr2018comprehensive,jia2019memguard}.

    \mypara{MNIST.} This is a dataset of 70,000 handwritten digits. The size of each image is 32 $\times$ 32. The images are cropped to 28 $\times$ 28 so that the digits are located at the center of the image. There are 10 classes of different digits in this dataset. There are 60000 training images and 10000 testing images in this dataset. This dataset is used to evaluate membership inference attack in ~\cite{shokri2017membership,salem2018ml}.
    
    \mypara{Target classifiers.} We choose the state-of-the-art NN architectures for these datasets, and use Pytorch~\cite{paszke2017automatic}, a widely used deep learning framework in academia, to implement these neural networks.

     
    
    For CIFAR-10 and CIFAR-100, we use AlexNet~\cite{krizhevsky2012imagenet}, VGG~\cite{Simonyan15}, 2 ResNet~\cite{He_2016_CVPR} architectures, and 1 DenseNet~\cite{Huang_2017_CVPR} architectures.  
    For PURCHASE-100 and TEXAS-100, we follow the NN architecture described in \cite{nasr2018comprehensive}. For the MNIST dataset, we follow the NN architecture described in \cite{shokri2017membership}.
    

    In the existing literature on MI attacks (e.g.,~\cite{shokri2017membership,nasr2018machine}), target classifiers are trained with a subset of the training instances.   More specifically, $10,000$ instances are used for CIFAR-10 dataset, CIFAR-100 dataset, TEXAS-100 dataset and MNIST dataset, and $20,000$ are used for PURCHASE-100, which contains about three times more data as the other datasets. 
    Part of the reason is that one needs instances not used in training to build shadow models. Also, as the training set size increases, a model's vulnerability to MI attacks becomes small.  As our purpose is to compare the effectiveness of different MI attacks and defenses, we use the same setting in most experiments in this paper.  We do study the effect of varying the number of training instances in Section~\ref{sec:vary_num}.

    The training recipe for all datasets and classifiers are summarized in Table~\ref{training_recipe} on Page \pageref{sec:appendix} in the appendix.  For widely-used benchmark datasets, namely CIFAR-10, CIFAR-100, and MNIST, the training and testing accuracy numbers in Table~\ref{training_recipe} are for the situation of using the whole training set for training.  As a result, these testing accuracy numbers in Table~\ref{training_recipe} are generally higher than those in membership inference experiments where classifiers are trained using only a subset of the instances.  These numbers are similar to the ones reported elsewhere, demonstrating that the training was done correctly. PURCHASE-100 dataset and TEXAS-100 dataset are not widely used benchmark datasets, and we report numbers when using $20,000$ and $10,000$ training instances, respectively.

\mypara{Using Shadow Models.} 
    Several MI attacks need to utilize ``shadow classifiers'' to train the attack classifiers.  We train $50$ shadow classifiers as our default choice and we follow the workflow proposed in \cite{shokri2017membership} with minor differences in generating the training sets for shadow classifiers and target classifiers, in order to achieve better efficiency for instance level attacks (the Instance-Vector attack). The target models and the shadow models are built in the exactly same way.

    We want the evaluation settings to reflect the assumptions that the adversary knows the general data distribution used in training, but not the exact training set. More specifically, we randomly divide $D$ into three disjoint parts: $D_{E}$, $D_{G}$ and $D_{H}$.  
    $D_{E}$ is the evaluation set and has 5000 instances.  MI attacks are evaluated on the accuracy of determining membership of instances in $D_E$.  
    The target classifier is trained with a randomly sampled set consisting of half of the instances in $D_{E}$ and the rest of training samples are sampled from $D_{G}$.  
    Each shadow classifier is trained with a randomly sampled half of the instances in $D_{E}$ and the rest of training samples are sampled from $D_{H}$.  This way, each shadow classifier is trained with half of the evaluation set, like the target classifier.  At the same time, at least $3/4$ of the instances used for training the target classifier are never used in training any shadow classifier.  

\subsection{Results for Existing Attacks}\label{sec:attack_results}

    An MI attack is itself a binary classification, and metrics such as false positive and false negative rates, F1, etc., will depend on the distribution of what percentage of the evaluation set consists of members.  Since there is no natural evaluation distribution, the approach taken in the literature~\cite{shokri2017membership,long2017towards,yeom2018privacy,nasr2018comprehensive} on MI attacks is to measure its accuracy over a balanced set that consists of half members and half non-members.  This measures an adversary's MI attack accuracy with a priori that the attacker does not favor a guess of member or non-member.  
    Since a trivial attack of randomly guessing whether an instance is a member has an accuracy of $1/2$, we define the {\em advantage} of an attack to be its accuracy over a balanced set minus $1/2$.


Figure~\ref{disjoint} shows the attack advantage of each attack on different classifier and dataset combinations. Table~\ref{disjoint_avg_attack_advantage} provides an average summary for all the attacks, including average training and testing accuracy.  


From Table~\ref{disjoint_avg_attack_advantage}, we can see that on MNIST the generalization gap is very small (around $0.02$), and the highest attack advantage among all attacks is also around $0.02$.  Essentially models on MNIST are almost immune to existing MI attacks, because it has a low generation gap. We thus do not include MNIST dataset in Figure~\ref{disjoint} or the following discussion of different attacks.  



\mypara{The Baseline attack.}  From Figure~\ref{disjoint}, we see that the empirically observed baseline attack advantage is very close to the theoretical prediction of half of the generalization gap.  We also observe that, comparing with other attacks, the baseline attack is pretty strong, even though it is trivial to implement and requires minimal access to the target classifier.  For example, on CIFAR-100 and TEXAS-100, the two datasets that are more vulnerable to MI attacks, the baseline attack can achieve $92.8\%$ and $74.9\%$ of the highest attack advantage, respectively.



\mypara{Comparing existing attacks.} From Figure \ref{disjoint} and Table \ref{disjoint_avg_attack_advantage}, we can observe that the Global-Loss attack and the Global-Probability attack can provide the highest attack advantage. The Class-Vector attack performs slightly worse than the highest two attacks. 

Based on the feature each existing attack uses, we can draw a conclusion that the probability of the correct label, which is used by the Global-Loss attack and the Global-Probability attack, is the most effective feature to launch MI attacks. Even though the Class-Vector attack can utilize the full prediction vector, it turns out that extra information does not help to improve the attack advantage. If we compare the Global-Probability attack with the Global-TopOne attack, we can see that the Global-TopOne attack performs strictly worse than the Global-Probability attack.  In addition, the differences in terms of advantage between these two attacks are highly correlated with training accuracy. This is because when training accuracy is low, the top-one probability for training many instances are not that of the correct label.  This shows that the probability of the correct label is better than the highest probability, in terms of attack advantage that can be achieved. The Global-TopThree attack provides similar performance as the Global-TopOne attack.

As for the Instance-Vector attack, since this attack is an instance-level attack, this attack requires significantly more shadow models than other existing attacks, meaning that the computational overhead is huge. In terms of attack advantage, this attack performs strictly worse than the Class-Vector attack, even though the input feature is the same, i.e. the full prediction vector. On CIFAR-100 dataset, this attack performs even worse than the Baseline attack. One possible reason is the choice of using the full prediction vector. Based on previous observation, if this attack uses only the probability of the correct label, it might achieve a higher attack advantage. Other possible reasons are the choice of KL-divergence metric and the number of shadow models.

In summary, we observe that simply using the probability of the correct label can achieve the highest attack advantage.  
In terms of constructing new defenses, one should aim to ensure that the distributions of the probability of the correct label for members and non-members are as similar as possible, which should also have the side-effect of reducing the generalization gap.  In this way, the attacks which rely on the probability of the correct label can be defended.

%% file: 4a_defenses.tex
\section{Existing Defenses and Our Proposed Defenses}\label{sec:defense}

\subsection{Existing Defenses}\label{sec:defense:existing}

Multiple defense mechanisms have been proposed in the existing literature to mitigate the threat of membership inference attacks. We summarize existing defenses in the following.

\mypara{Reducing generalization gap by increasing test accuracy.}
In ~\cite{shokri2017membership}, Shokri et al.~showed that using $L_2$-Regularizer, which could reduce overfitting, can help defend against MI attacks.  
Salem et al.~\cite{salem2018ml} explored using Dropout, which was first proposed in ~\cite{srivastava2014dropout} to reduce overfitting, as a defense against MI attacks.  As these approaches improve testing accuracy and thus reduces the generalization gap, they indeed reduce vulnerability to MI attacks.  However, in many cases, these approaches are already deployed, and the generalization gap (and hence vulnerability to MI attacks) remain significant.  Additional defenses are needed for these cases.

\mypara{Hiding prediction information.}
In ~\cite{shokri2017membership}, it was proposed to reduce the information given by the prediction vectors, such as providing only the top-k probabilities and using high temperature in softmax.  This has limited effectiveness as the top-k probabilities give enough information needed by the best attacks, and high temperatures change only the absolute values of confidences, but not the fact that confidences for members tend to be higher than that for non-members.

\mypara{Model Stacking and Adversary-Specific Defense.}
Salem et al.~\cite{salem2018ml} also proposed to use model stacking as a defense. Model stacking is a common ensemble technique used in machine learning applications, which combines multiple weak classifiers together to make the final prediction.

Nasr et al.~\cite{nasr2018machine} proposed a \textbf{Min-Max Game} style defense to train a secure target classifier. During the training of the target classifier, a defender's attack classifier is trained simultaneously to launch the membership inference attack. The optimization objective of the target classifier is to reduce the prediction loss while minimizing the membership inference attack accuracy. This is equivalent to adding a new regularization term to the training process, which is called \textit{adversarial regularization}.  

Jia et al.~\cite{jia2019memguard} proposed the \textbf{Mem-Guard} defense.  In this defense, one trains an MI attack model in addition to the target classifier.  When the target classifier is queried with an instance, the resulting prediction vector is not directly returned.  Instead, one tries to find a perturbed version of the vector such that the perturbation is minimal, the prediction label is not changed, and the MI attack model output (0.5,0.5) as its prediction vector.  This perturbed vector is then returned. In~\cite{jia2019memguard}, it is shown that Mem-Guard is more effective than Min-Max Game and Model Stacking.  We compare with Mem-Guard in Section~\ref{sec:diff_regu_eval}.



\mypara{Knowledge Distillation.}
Distillation was proposed in ~\cite{hinton2015distilling} for the purpose of model compression, and used in~\cite{shejwalkar2019reconciling} as a defense against MI attacks.
In~\cite{shejwalkar2019reconciling}, one first trains a teacher model using the private training set, and then trains a student model using another unlabeled dataset from the same distribution as the private set.  The training objective is to minimize the KL-divergence between the student model's output and the probability vector predicted by the teacher model under a pre-set temperature $T$ in softmax function.  The student model is given as the output.  The intuition is that since the student model is not directly optimized over the private set, their membership may be protected.  In ~\cite{kaya2020effectiveness}, the authors compared the distillation technique with other regularization techniques.  However, the most effective attacks were not considered in the evaluation in~\cite{kaya2020effectiveness,shejwalkar2019reconciling}.  Also, membership of the second (unlabelled) dataset is not considered.  Thus this technique can be applied only when there exists a public dataset for training. 

\mypara{Differential Privacy.}
Differential privacy ~\cite{dwork2008differential,dwork2006calibrating,Dwo06} is a widely used privacy-preserving technique.  A differential privacy based defense technique adds noise to the training process of a classifier.  One example is DP-SGD proposed in ~\cite{abadi2016deep}. Noise is added to the gradient to ensure data privacy.  Training with differential privacy provides theoretical guarantee against any MI attacks. 
 
Currently, achieving a meaningful theoretical guarantee (e.g., with a resulting $\epsilon < 5$) requires the usage of very large noises.  However, one could use much smaller noises in DP-SGD.  While doing this fails to provide a meaningful theoretical guarantee (the $\epsilon$ value would be too large), adding noises in the training process can nonetheless provide empirical defense against MI attacks. In this paper, we conduct an empirical comparison with DP-SGD to see which defenses achieve a better tradeoff between testing accuracy and resiliency against MI attacks.  

\begin{figure*}
     \centering
     \includegraphics[width=18cm,height=9cm]{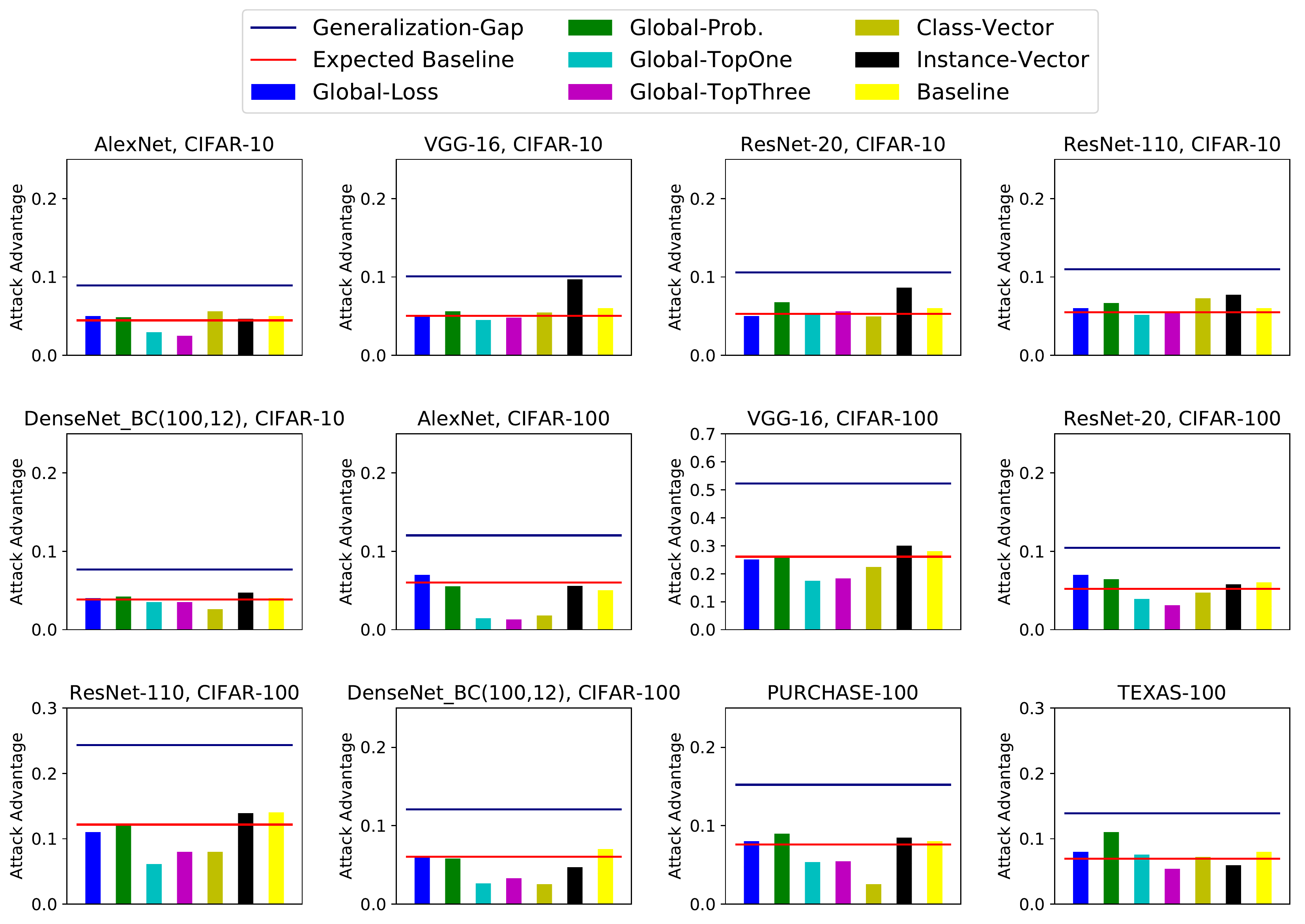}
     \caption{Attack advantages of different attacks after defense for 5 models each on CIFAR-10 and CIFAR-100, and 1 model each for PURCHASE-100 and TEXAS-100. Each bar represents a different attack. The higher horizontal line represents the generalization gap $g$ and the lower horizontal line represents the expected baseline attack advantage ($g/2$).}
     \label{fig:defense_disjoint}
 \end{figure*}
 
 \begin{table*}
     \normalsize
     \centering
     \begin{center}
    \begin{tabular}{lrrrrrrrrr}
        
        \multicolumn{1}{c}{Dataset-Model}  & \multicolumn{3}{c}{Testing Accuracy} & \multicolumn{3}{c}{Highest Attack Advantage} & \multicolumn{3}{c}{Generalization Gap} \\
        \hline 
        \\
        \multicolumn{1}{c}{-} & w/o def. & w/ def. &\multicolumn{1}{c}{difference} & w/o def. & w/ def.  &  \multicolumn{1}{c}{drop}  & w/o def. & w/ def. & \multicolumn{1}{c}{drop} \\
        
        \cmidrule(lr){1-1}
        \cmidrule(lr){2-4}
        \cmidrule(lr){5-7}
        \cmidrule(lr){8-10}

        \multicolumn{1}{c}{CIFAR-10 - Alexnet} & \multicolumn{1}{c}{0.652} & \multicolumn{1}{c}{0.662}  & \multicolumn{1}{c}{\textbf{+0.01}} & \multicolumn{1}{c}{0.210} & \multicolumn{1}{c}{0.056}  & \multicolumn{1}{c}{\textbf{73.3\%}} & \multicolumn{1}{c}{0.337} & \multicolumn{1}{c}{0.089}  & \multicolumn{1}{c}{\textbf{73.6\%}} \\

        \multicolumn{1}{c}{CIFAR-10 - VGG-16} & \multicolumn{1}{c}{0.811} & \multicolumn{1}{c}{0.809} & \multicolumn{1}{c}{\textbf{-0.002}} & \multicolumn{1}{c}{0.162} & \multicolumn{1}{c}{0.097}  & \multicolumn{1}{c}{\textbf{40.1\%}} & \multicolumn{1}{c}{0.169} & \multicolumn{1}{c}{0.108}  & \multicolumn{1}{c}{\textbf{36.1\%}} \\

        \multicolumn{1}{c}{CIFAR-10 - ResNet-20} & \multicolumn{1}{c}{0.820} & \multicolumn{1}{c}{0.818}  & \multicolumn{1}{c}{\textbf{-0.002}}& \multicolumn{1}{c}{0.130} & \multicolumn{1}{c}{0.086}  & \multicolumn{1}{c}{\textbf{33.8\%}} & \multicolumn{1}{c}{0.171} & \multicolumn{1}{c}{0.106}  & \multicolumn{1}{c}{\textbf{38.1\%}} \\

        \multicolumn{1}{c}{CIFAR-10 - ResNet-110} & \multicolumn{1}{c}{0.794} & \multicolumn{1}{c}{0.784}  & \multicolumn{1}{c}{\textbf{-0.01}}& \multicolumn{1}{c}{0.150} & \multicolumn{1}{c}{0.077}  & \multicolumn{1}{c}{\textbf{48.7\%}} & \multicolumn{1}{c}{0.201} & \multicolumn{1}{c}{0.110}  & \multicolumn{1}{c}{\textbf{45.3\%}} \\

        \multicolumn{1}{c}{CIFAR-10 - DenseNet-BC(100,12) } & \multicolumn{1}{c}{0.801} & \multicolumn{1}{c}{0.802} & \multicolumn{1}{c}{\textbf{+0.001}} & \multicolumn{1}{c}{0.137} & \multicolumn{1}{c}{0.047}  & \multicolumn{1}{c}{\textbf{65.7\%}} & \multicolumn{1}{c}{0.196} & \multicolumn{1}{c}{0.077}  & \multicolumn{1}{c}{\textbf{60.7\%}} \\
        
        \multicolumn{1}{c}{\textbf{CIFAR-10 - Average of 5 models}} & \multicolumn{1}{c}{\textbf{0.776}} & \multicolumn{1}{c}{\textbf{0.775}} & \multicolumn{1}{c}{\textbf{-0.001}} & \multicolumn{1}{c}{\textbf{0.158}} & \multicolumn{1}{c}{\textbf{0.073}}  & \multicolumn{1}{c}{\textbf{53.8\%}} & \multicolumn{1}{c}{\textbf{0.215}} & \multicolumn{1}{c}{\textbf{0.098}}  & \multicolumn{1}{c}{\textbf{54.4\%}} \\
        
        \hline

        \multicolumn{1}{c}{CIFAR-100 - Alexnet} & \multicolumn{1}{c}{0.248} & \multicolumn{1}{c}{0.238} & \multicolumn{1}{c}{\textbf{-0.01}} & \multicolumn{1}{c}{0.284} & \multicolumn{1}{c}{0.071}  & \multicolumn{1}{c}{\textbf{75.0\%}} & \multicolumn{1}{c}{0.751} & \multicolumn{1}{c}{0.121}  & \multicolumn{1}{c}{\textbf{83.9\%}} \\

        \multicolumn{1}{c}{CIFAR-100 - VGG-16} & \multicolumn{1}{c}{0.285} & \multicolumn{1}{c}{0.275}  & \multicolumn{1}{c}{\textbf{-0.01}}& \multicolumn{1}{c}{0.411} & \multicolumn{1}{c}{0.300}  & \multicolumn{1}{c}{\textbf{27.0\%}} & \multicolumn{1}{c}{0.715} & \multicolumn{1}{c}{0.542}  & \multicolumn{1}{c}{\textbf{24.2\%}} \\

        \multicolumn{1}{c}{CIFAR-100 - ResNet-20} & \multicolumn{1}{c}{0.446} & \multicolumn{1}{c}{0.456} & \multicolumn{1}{c}{\textbf{+0.01}} & \multicolumn{1}{c}{0.295} & \multicolumn{1}{c}{0.073}  & \multicolumn{1}{c}{\textbf{75.3\%}} & \multicolumn{1}{c}{0.542} & \multicolumn{1}{c}{0.127}  & \multicolumn{1}{c}{\textbf{76.6\%}} \\

        \multicolumn{1}{c}{CIFAR-100 - ResNet-110} & \multicolumn{1}{c}{0.424} & \multicolumn{1}{c}{0.426}  & \multicolumn{1}{c}{\textbf{+0.002}}& \multicolumn{1}{c}{0.342} & \multicolumn{1}{c}{0.145}  & \multicolumn{1}{c}{\textbf{57.6\%}} & \multicolumn{1}{c}{0.574} & \multicolumn{1}{c}{0.243}  & \multicolumn{1}{c}{\textbf{57.7\%}}\\

        \multicolumn{1}{c}{CIFAR-100 - DenseNet-BC(100,12) } & \multicolumn{1}{c}{0.428} & \multicolumn{1}{c}{0.430} & \multicolumn{1}{c}{\textbf{+0.002}} & \multicolumn{1}{c}{0.256} & \multicolumn{1}{c}{0.074}  & \multicolumn{1}{c}{\textbf{71.1\%}} & \multicolumn{1}{c}{0.511} & \multicolumn{1}{c}{0.121}  & \multicolumn{1}{c}{\textbf{76.3\%}} \\
        
        \multicolumn{1}{c}{\textbf{CIFAR-100 - Average of 5 models}} & \multicolumn{1}{c}{\textbf{0.366}} & \multicolumn{1}{c}{\textbf{0.365}} & \multicolumn{1}{c}{\textbf{-0.001}} & \multicolumn{1}{c}{\textbf{0.318}} & \multicolumn{1}{c}{\textbf{0.133}}  & \multicolumn{1}{c}{\textbf{58.2\%}} & \multicolumn{1}{c}{\textbf{0.619}} & \multicolumn{1}{c}{\textbf{0.231}}  & \multicolumn{1}{c}{\textbf{62.9\%}} \\
        
        \hline
        
        \multicolumn{1}{c}{PURCHASE-100 - PURCHASE-100} & \multicolumn{1}{c}{0.791} & \multicolumn{1}{c}{0.781}  & \multicolumn{1}{c}{\textbf{-0.01}}& \multicolumn{1}{c}{0.142} & \multicolumn{1}{c}{0.089}  & \multicolumn{1}{c}{\textbf{37.3\%}} & \multicolumn{1}{c}{0.191} & \multicolumn{1}{c}{0.152}  & \multicolumn{1}{c}{\textbf{20.4\%}}\\

        \multicolumn{1}{c}{TEXAS-100 - TEXAS-100} & \multicolumn{1}{c}{0.535} & \multicolumn{1}{c}{0.539}  & \multicolumn{1}{c}{\textbf{+0.004}}& \multicolumn{1}{c}{0.191} & \multicolumn{1}{c}{0.110}  & \multicolumn{1}{c}{\textbf{42.4\%}} & \multicolumn{1}{c}{0.280} & \multicolumn{1}{c}{0.139}  & \multicolumn{1}{c}{\textbf{50.4\%}} \\

       \bottomrule
    \end{tabular}
    \end{center}
     \caption{The testing accuracy, the highest attack advantage and the generalization gap after applying our defense under various settings. The percentage of drop is relative drop. The maximum absolute testing accuracy difference is limited by $1\%$.}
     \label{tab:defense_summary}
\end{table*}

\subsection{Our proposed defenses}

Our experimental results show that the generalization gap is directly related to vulnerability to MI attacks.  Thus the first line of defense against MI attacks is to reduce overfitting, which is also one of the primary goals of ML research itself.  
Our experimental results also show that the feature that enables the most effective MI attacks is the confidence level of the true label for an instance.  We call this the label confidence of an instance.  MI attacks exploit the fact that members, in general, have higher label confidence values than non-members do.  Thus a promising idea for defenses is to make the distribution of label confidence for members to be close to that of non-members. 
We propose a defense that combines the above two ideas. 


\mypara{Mix-up Training Augmentation.}  
The first piece of our proposed defense is mix-up training, which was first introduced by Zhang et al.~\cite{zhang2018mixup}.  Mix-up training uses linear interpolation of two different training instances to generate a mixed instance and train the classifier with the mixed instance. The generation of mixed instances can be described as follows:
        \begin{align}
        \tilde{x} & = \lambda x_i + (1-\lambda) x_j, \nonumber\\ 
        \tilde{y} & = \lambda y_i + (1-\lambda) y_j, \label{eq:mixup}
    \end{align}
    where $\lambda \sim beta(\alpha,\alpha), \alpha \in (0,\infty).$
    Here $x_i$ and $x_j$ in equation \ref{eq:mixup} are instance feature vectors randomly drawn from the training set; $y_i$ and $y_j$ are one-hot label encodings corresponding to $x_i$ and $x_j$. $(\tilde{x},\tilde{y})$ is used in training.
In Zhang et al.~\cite{zhang2018mixup} it is shown that {\em mix-up training} can improve generalization, resulting in higher accuracy on CIFAR-10 and CIFAR-100.  This, in turn, reduces the generalization gap.  Also, intuitively, since only the mixed instances are used in training, the classifier will not be directly trained in the original training instances, and should not remember them as well.  

\mypara{MMD-based Regularization.}
The second piece of our proposed defense aims at making the distribution of label confidence for members to be close to that of non-members. 
This is achieved 
by adding to the training loss function a regularizing term that is based on the difference between the probability vector distribution of the minibatch in training, and that from a validation set.  

We need a differentiable metric to measure the difference, and choose to use Maximum Mean Discrepancy (MMD)~\cite{fortet1953convergence,gretton2012kernel}.  MMD is used to construct statistical tests to determine if two samples are drawn from different distributions, based on Reproducing Kernel Hilbert Space (RKHS)~\cite{borgwardt2006integrating}. Let $X = (x_1,\cdots,x_n)$ and $Y = (y_1,\cdots,y_m)$ be the random variable sets drawn from distribution $\mathcal{P}$ and $\mathcal{Q}$. The empirical estimation of distance between $\mathcal{P}$ and $\mathcal{Q}$, as defined by MMD, is:
    \begin{equation}\label{eq:MMD}
        \text{Distance}(X,Y) = \norm{\frac{1}{n}\sum_{i=1}^n\phi(x_i)-\frac{1}{m}\sum_{j=1}^m\phi(y_j)}_\mathcal{H},
    \end{equation}
where $\mathcal{H}$ is a universal RKHS, and $\phi : \mathcal{X} \mapsto \mathcal{H}$ is a function mapping each sample to a point in $\mathcal{H}$.  
In MMD, $\phi$ should be chosen to be the function that maximizes the resulting norm, and is usually modeled using a neural network with the parameters trained~\cite{long2017deep}.  
In our context, each $x_i$ ($y_i$) is the softmax output of the $i$-th training (validation) instance. 
To reduce training time, we use a more traditional approach and make $\phi$ the Gaussian kernel~\cite{gretton2012kernel}.
The sums $\sum_{i=1}^n\phi(x_i)$ and $\sum_{j=1}^m\phi(y_j)$ are empirical estimates of the mean embeddings of the training and validation softmax distributions~\cite{gretton2012kernel} and, thus, Equation (\ref{eq:MMD}) is a valid two-sample test. We adopt the implementation from ~\cite{transferlearning.xyz}. We leave the task of performing a stronger two-sample test with a minimax-optimized MMD objective as future work.
Note that the regularizer Distance in Equation (\ref{eq:MMD}) is invariant to permutations of the ordering of the training ($x_1,\ldots,x_n$) and validation ($y_1,\ldots,y_n$) softmax outputs of the data instances, hence, Distance is a permutation-invariant function (a set function).
   
    
 In order to calculate the MMD regularization loss, we need one mini-batch of training and validation instances. In the implementation of our MMD loss, we choose to reduce the difference between the probability vector distributions of members and non-members for the same class. That is, a batch of training samples and a batch of validation samples in the same class are used together to compute the MMD score. Since MMD is differentiable, we obtain two sets of gradients: the first set is calculated based on training instances; the second set is based on validation instances. To update our classifier, we only use the first set of gradients, since the second set of gradients would optimize over the validation instances, which is undesirable as it could overfit them and make the empirical distribution $(y_1,\ldots,y_m)$ significantly different from that of the (future) test data. 
 
 In the experiments, we ensure that the validation set is disjoint from the training set of any target or shadow models and the evaluation set for MI attacks. The size of the training set and the validation set are the same. 

 Since technically the validation set is part of the input for the training process, one natural question is whether instances in the validation set would also be vulnerable to MI attacks.  However, since we discard the set of gradients with respect to the validation set in training, the model is not directly optimized to fit the distribution of the validation set.  Thus it is expected that the MI attacks against the validation set are not effective. This is verified in experiments that will be reported in section \ref{sec:MI_validation}.

    
    

%% file: 4b_defense_eval.tex
\section{Evaluation of Proposed Defense} \label{sec:eval}

We have already described the detailed experimental settings in Section \ref{sec:experimental_setup_1}, and summarized the evaluation results of the MI attacks in Section \ref{sec:attack_results}. 
Here we evaluate the effectiveness of our proposed defenses.

\subsection{Effectiveness of Proposed Defenses}\label{sec:defense_results}


We now show that our proposed defenses are effective at reducing the generalization gap and defending against MI attacks.  When using the MMD defense, one can choose the weight of MMD loss relative to training loss.  As a larger weight results in a lower generalization gap but worse testing accuracy, the choice of this weight results in different trade-offs between privacy and accuracy.   For the experiments in this subsection, we choose the largest weight while ensuring that the absolute testing accuracy difference is less than $1\%$ when compared to the classifiers trained without defense.  The weight parameters used for different datasets can be found in Table \ref{training_recipe} on Page~\pageref{training_recipe}.  In Section~\ref{sec:diff_regu_eval}, we show the results of varying the weight and compare that to other defenses. 


Figure \ref{fig:defense_disjoint}  shows the attack advantages of different attacks in various settings when using our proposed defense.  For model architectures on CIFAR-10, CIFAR-100 and TEXAS-100, we use MMD \& mix-up.  For PURCHASE-100, using mix-up training leads to a significant accuracy drop, which may be due to the fact that the data features are binary. We thus use only MMD training for them. 

Table \ref{tab:defense_summary} summarizes the information in Figure \ref{fig:defense_disjoint} by showing the highest attack advantage for each setting.  We can see that \textbf{our defense can significantly reduce the effectiveness of existing attacks while keeping the testing accuracy.}  The testing accuracy drop is less than $1\%$ and the highest attack advantage is significantly reduced. For CIFAR-10, the average highest attack advantage is reduced from $ 0.158 $ to $ 0.073 $ ($ 53.8\% $ drop). For CIFAR-100, the average highest attack advantage drop is reduced from $ 0.318 $ to $ 0.133 $ ($ 58.2\% $ drop).  For TEXAS-100, the highest attack advantage drops $42.4\%$. For PURCHASE-100, with only the MMD-Loss defense, the highest attack advantage drops $37.3\%$.

From Table \ref{tab:defense_summary} we can also see that \textbf{the generalization gap is significantly reduced} for CIFAR-10 and CIFAR-100. For CIFAR-10, the average generalization gap is reduced from $0.215$ to $0.098$ ($54.4\%$ drop). For CIFAR-100, the average generalization gap is reduced from $0.619$ to $0.231$ ($62.7\%$ drop). For TEXAS-100, the generalization gap drops $50.4\%$. For PURCHASE-100, with only the MMD-Loss defense, the generalization gap drops $20.4\%$.

\subsection{Comparison with Other Defenses}\label{sec:diff_regu_eval}

\begin{figure*}
     \centering
     \includegraphics[width=18cm,height=10cm]{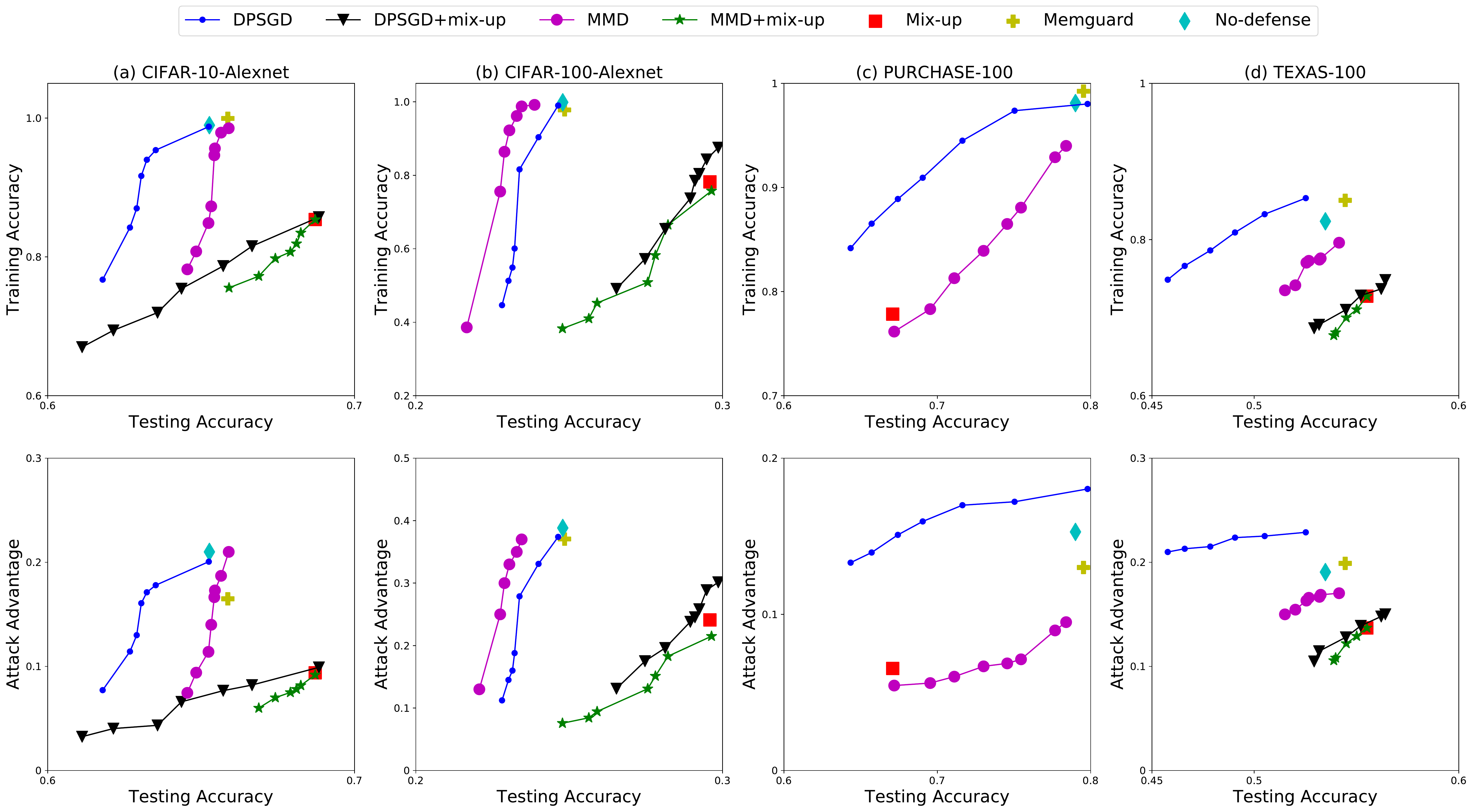}
     \caption{Comparison of different defense techniques on CIFAR-10, CIFAR-100, PURCHASE-100 and TEXAS-100.}
     \label{fig:diff_regu_plot}
 \end{figure*}

We compare our proposed defenses with the best defenses among the existing defenses that are described in Section~\ref{sec:defense:existing}. In the experiments, we assume that the attackers are aware of the applied defense and are able to train shadow models with the same defense. In modern deep neural network training, $L_2$-regularizer and dropout are widely used. In our experiment, we use these two techniques in all approaches, including the baseline model (which we call no-defense). 
Model stacking, according to~\cite{jia2019memguard}, will incur high label loss, which implies high testing accuracy drop.  Also according to~\cite{jia2019memguard}, the Mem-Guard defense outperforms MIN-MAX Game defense~\cite{nasr2018machine}.  
We found that knowledge distillation results in either training failure or high accuracy drop (see Appendix~\ref{sec:knowledge_distillation} for details); thus we do not include data on knowledge distillation here. 


Figure \ref{fig:diff_regu_plot} shows the results comparing our proposed defense with the state-of-the-art empirical defense Mem-Guard and DP-SGD.  We also notice that our proposed mix-up training is compatible with DP-SGD, thus we also evaluate DP-SGD+mix-up for CIFAR-10, CIFAR-100 and TEXAS-100.
DP-SGD, DP-SGD+mix-up, MMD+Mix-up, and MMD-only have tunable parameters and are plotted as curves.  The other methods do not have tunable parameters and show up as a single point in the graph.  The first row in Figure~\ref{fig:diff_regu_plot} plots testing accuracy against training accuracy, and the second row plots testing accuracy against the highest attack advantage. 
In all figures, bottom-right means high testing accuracy and low generalization gap/attack advantage, and is thus preferred.  


\textbf{Mix-up versus Mem-Guard and No-defense.}
Comparing Mix-up with Mem-Guard and No-Defense in Figure~\ref{fig:diff_regu_plot}, we can see that on CIFAR-10, CIFAR-100, and TEXAS-100, Mix-up increases testing accuracy while decreasing training accuracy. Mix-up also provides lower attack advantages comparing to Mem-Guard and No-defense. However, on PURCHASE-100, Mix-up results in a significant drop in accuracy. 

\textbf{MMD+Mix-up vs. DP-SGD+Mix-up.}
On CIFAR-10, CIFAR-100, and TEXAS-100 we use DP-SGD+Mix-up and MMD+Mix-up. we observe that MMD+Mix-up generally provide lower training accuracy and lower attack advantage for the same testing accuracy. 
Interestingly, we observe that on CIFAR-100 and TEXAS-100, DP-SGD+Mixup sometimes provides slightly higher (by around $1\%$) testing accuracy than using Mixup alone.  
Looking at the experimental data, we found that when the noise scale falls into the range from $1e-8$ to $5e-8$, applying DP-SGD would increase the testing accuracy.  It appears that adding slight random perturbation to the gradients improves the generalization when Mix-up is used.  
This improvement occurs only when the noise scale is in the range.  When the noise is too large, testing accuracy drops.  When the noise is too small, the impact is too small to affect the trained model.  We did not observe the same phenomenon for DP-SGD without Mix-up.  
However, we note that when accuracy is improved, the highest attack advantage is also improved and these models are still vulnerable to MI attacks.  If we want to reduce the MI vulnerability, DP-SGD+Mix-up offers lower accuracy than MMD+Mix-up under the same level of MI vulnerability. 

\textbf{Conclusion.}  When Mix-up results in an improvement in testing accuracy, MMD+mix-up provides the best defense. When Mix-up does not result in accuracy improvement (as in the PURCHASE-100 dataset), MMD provides the best defense.

\begin{table*}[]
    \centering
    \vspace{-0.5cm}
    
    \begin{tabular}{l r r r r}
        \toprule

        \multicolumn{1}{c}{Dataset} &  \multicolumn{1}{c}{Testing Accuracy} &  \multicolumn{1}{c}{Validation Accuracy} &  \multicolumn{1}{c}{Highest Attack Advantage} \\
        
        \cmidrule(lr){1-1}
        \cmidrule(lr){2-2}
        \cmidrule(lr){3-3}
        \cmidrule(lr){4-4}
        
         \multicolumn{1}{c}{CIFAR-10 - AlexNet} & \multicolumn{1}{c}{0.662}& \multicolumn{1}{c}{0.661} &  \multicolumn{1}{c}{0.005}\\
         
         \multicolumn{1}{c}{CIFAR-100 - AlexNet} & \multicolumn{1}{c}{0.238} &\multicolumn{1}{c}{0.235} &  \multicolumn{1}{c}{0.002}\\
         
         \multicolumn{1}{c}{PURCHASE-100} &  \multicolumn{1}{c}{0.781}&  \multicolumn{1}{c}{0.773} &   \multicolumn{1}{c}{0.007}\\
         
         \multicolumn{1}{c}{TEXAS-100} & \multicolumn{1}{c}{0.539} &\multicolumn{1}{c}{0.541} & \multicolumn{1}{c}{0.003} \\

        \bottomrule
    \end{tabular}
    \caption{Validation accuracy and testing accuracy of models trained with our MMD+mix-up defense is similar. MI attacks are no more effective than random guessing on evaluation set which consists one half validation data and one half testing data.}
    \label{tab:validation_MI}
\end{table*}

\subsection{Membership inference on validation set}\label{sec:MI_validation}

Training with MMD loss requiring using a validation set.  Since the training process uses both the training set and the validation set, one question is whether membership of the validation set will be leaked.  Since the model is not trained to minimize the loss on the validation set, we do not expect the model to leak membership information for instances in the validation set.  

To verify whether our expectation is correct, we apply MI attacks to a balanced evaluation set which consists one half validation data (labeled as members) and one half testing data (non-members). The attack models are trained using the same way as Section \ref{sec:defense_results}.

We summarize the results in Table \ref{tab:validation_MI}. The difference between validation accuracy and testing accuracy for models trained with our proposed defense is less than $1\%$.  
The highest attack advantage among all different attacks for all datasets is less than $1\%$, which indicates that there is no MI threat for instances in the validation set.

\begin{figure}[]
    \centering
    \includegraphics[width=6cm,height=6cm]{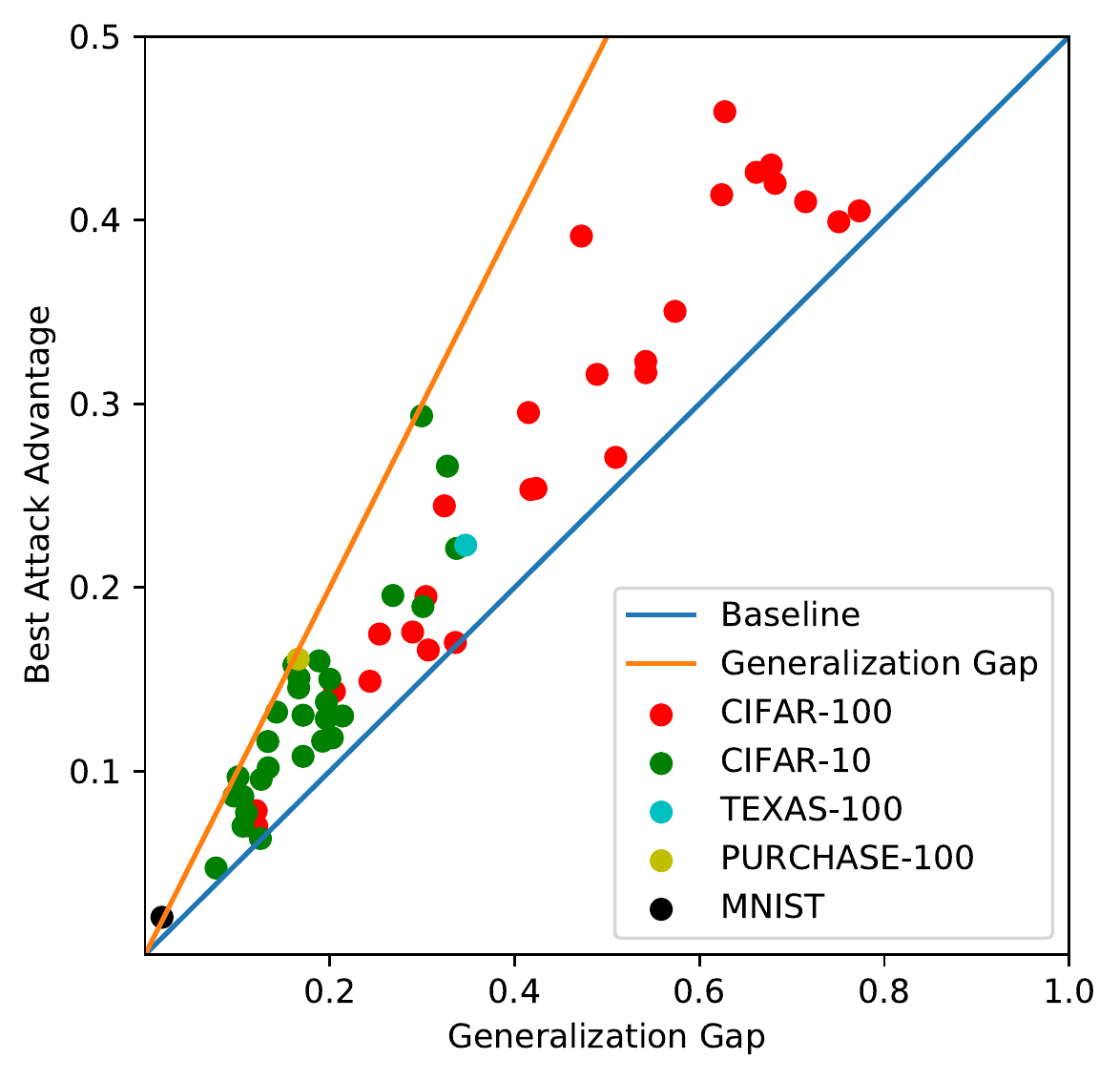}
    \caption{Generalization Gap vs Highest Attack Advantage.  Each point represents one classifier on one dataset.}
    \label{fig:gapvsbset}
\end{figure}

\subsection{Bound between Generalization Gap and the Highest attack advantage}

Based on our extensive evaluation of existing attacks on different datasets and models, we notice that the baseline attack could achieve decent attack advantages comparing to the best one of existing attacks, before and after defenses.
Meanwhile, we also observe an interesting phenomenon: the generalization gap $g$ is an empirically upperbound for the best attack advantage, as shown in Figure \ref{fig:gapvsbset}. On the scatter plot, each point represents one classifier on one dataset.  The $x$ axis gives the generalization gap $g$, and the $y$ axis gives the best attack advantage $v$. We empirically observe that 
    \begin{align}
        \frac{g}{2} & \le v \le g  \label{eq:one}
    \end{align}
We leave additional analysis and validation of this phenomenon for future work.

%% file: 6_related_work.tex
\section{Related Work} \label{sec:related}


    
    Membership inference attacks from summary statistics were studied in the context of Genome-Wide Association Studies (GWAS).  
    Homer et al.~\cite{homer2008resolving} proposed attacks that could tell with high confidence whether an individual participated in a GWAS study, assuming that the individual's DNA is known.  The attack works even if the group includes hundreds of individuals.  Because of the privacy concerns from such attacks, a number of institutions, including the US National Institute of Health (NIH) and the Welcome Trust in London all decided to restrict access to data from GWAS.
    Wang et al~\cite{WLW+09} improved these attack techniques. 

    Li et al.~\cite{LQS+13} introduced the membership privacy framework, which prevents an adversary from significantly improving its confidence that an entity is not in the dataset or not, and showed that $\epsilon$-differential privacy is equivalent to membership privacy where the adversary's prior belief is such that the probability of any instance's membership is independent from that of any other instance.  This framework was used in Backes et al.~\cite{backes2016membership} to address privacy concerns with genomics data. 
    
    Shokri et al.~\cite{shokri2017membership} presented the first study of MI attack on classification models.  The Class-Vector attack evaluated in this paper is from \cite{shokri2017membership}.  
    Shokri et al.~\cite{shokri2017membership} observed that overfitting is highly related to MI attacks, and regularization techniques such as dropout and L2 regularization may be applied to mitigate MI attacks.  It is shown in Shokri et al.~\cite{shokri2017membership}  that strong L2 regularization leads to a rapid drop in test accuracy.  The classifier models considered in this paper are already trained using dropout.  

    Yeom et al.~\cite{yeom2018privacy} quantitatively analyzed the relationship between the MI attacks and the loss over both training set and testing set.  The Global-Loss attack and the baseline attack are from this paper. 
    Salem et al.~\cite{salem2018ml} studied the effect of MI attacks using less and less information. 
    They reported that only one shadow model would be sufficient to launch the attack proposed in Shokri et al.~\cite{shokri2017membership}. 
    The Global-TopOne and Global-TopThree attacks are from this paper. 
    These attacks are evaluated in our paper.  See Section~\ref{sec:attack} for details on these attacks. 
    For defense, Salem et al.\cite{salem2018ml} proposed using Dropout and model stacking as defenses against MI attacks.  
    Since model stacking also reduces the generalization gap, it can also mitigate the threat of MI attacks.
    

    
    Long et al.\cite{long2017towards} proposed the Instance-Vector attack.  
    Long et al.~\cite{long2018understanding} first assess which instances are more vulnerable to MI attacks, and then carry out attacks against those vulnerable ones.  Similar attacks can be carried out by analyzing the distribution of an instance's predicted probability in shadow models, and are interesting future work. 
    
    Nasr et al.~\cite{nasr2018machine} proposed a privacy MIN-MAX game to help defense the MI attack. The main idea comes from GAN \cite{goodfellow2014generative}.  By replacing the discriminator with an attacker, the benign users can train the target model and the attack model simultaneously. The expected result is that a target model that can resist the attack used in the training.  From the numbers reported in \cite{nasr2018machine}, one can see that while models trained in this way can resist against MI attacks in Nasr et al.~\cite{nasr2018machine}, the generalization gap is such that the baseline attack would have higher advantages.  
     %
    Nasr et al.~\cite{nasr2018comprehensive} introduced white-box MI attacks which leveraged the gradient information. Their experiments showed that attacks exploiting the gradient corresponding to the last layer can achieve higher advantages than the generalization gap.  Investigating white-box attacks is interesting future work. 

    Truex~\cite{truextowards} studied MI attacks on traditional machine learning algorithms such as logistic regression and K-nearest neighbors. Their paper showed that logistic regression, K-nearest neighbors, decision tree and Naive Bayes can be easily attacked by the membership attack.  Carlini et al.~\cite{carlini2018text} investigated membership inference attacks on text data and showed that the deep language models can unintentionally memorize training data, which is vulnerable to membership inference attack.  
    Hayes et al.~\cite{hayes2019logan} showed that GAN is also vulnerable to membership attack. They utilized the output of the discriminator of the GAN to determine the membership.
    We point out that the baseline attack and attacks based on the predicted probability of the correct label can be carried out in these settings as well. 
    
    Jia et al.~\cite{jia2019memguard} brought adversarial example generation into defending membership inference attack. In their defense, which is called ``Mem-Guard'', the defender trains its own membership inference attack model and by generating adversarial example based on each instance to fool the defender's attack model, the defender can use the ``adversarial perturbation'' as the ``noise''. After adding the ``noise'' to the prediction vector of each instance, the noise will help to fool the attacker's attack model.
    
    shejwalkar et al.~\cite{shejwalkar2019reconciling} proposed to use knowledge distillation as a defense mechanism. The defender first trains a teacher model using a private dataset and then this teacher model is used to generate labels for an unlabeled validation set which is drawn from the same distribution as the private set. Another student model is trained using the validation set with the predicted labels. This student model is not directly optimized over the private set thus the membership information of the private set will not be significantly leaked when publishing the student model.
    
    Kaya et al. ~\cite{kaya2020effectiveness} evaluated multiple different regularization techniques, such as knowledge distillation, label smoothing and dropout. They show that the distillation, spatial dropout and data augmentation are the most effective ones regarding defeating MI attacks on CNN models.


%% file: 7_conclusion.tex
\section{Conclusions} \label{sec:conclusion}

In this work, we have conducted extensive experiments on evaluating existing MI attacks. We found that the baseline attack is a surprisingly strong attack, and the most effective feature for doing MI attacks is the probability of the correct label. By exploiting these findings, we propose a new MI defense that intentionally reduces the training accuracy as a way to defend against MI attacks.  
In particular, our defense is based on a new regularizer (MMD loss) and combines MMD loss with mix-up training whenever mix-up training improves accuracy.  
Our experiments show that our proposed defenses are able to maintain test accuracy while significantly reducing the model's vulnerability to MI attacks, and improves upon the state of the art in defenses against MI attacks.



A few open questions remain.  For instance, whether the generalization gap can be closed completely without reducing the test accuracy.  Another intriguing question is whether there is an explanation for the empirical observation that the generation gap seems to upper bound black-box MI attack advantages.  

%% file: 8_appendix.tex
\section{Appendix} \label{sec:appendix}

\subsection{Training recipe and model performance}
    We report the training accuracy and testing accuracy for each model when training from scratch. The training set size is 50000,50000,\\10000,20000 and 60000 for CIFAR-10 dataset, CIFAR-100 dataset, TEXAS-100 dataset, PURCHASE-100 dataset and MNIST dataset. See Table \ref{training_recipe} for more details.

\subsection{Evaluation of Knowledge Distillation}
\label{sec:knowledge_distillation}
We follow the experiment settings described in \ref{sec:experimental_setup_1} and we find that knowledge distillation using proposed parameters from~\cite{shejwalkar2019reconciling} result in training failure on CIFAR-100 using AlexNet. More specifically, the training accuracy and the testing accuracy of the privately trained model are both less than $3\%$. For PURCHASE-100 and TEXAS-100, after applying knowledge distillation, the training accuracy and the testing accuracy are both less than $3\%$. For AlexNet trained with knowledge distillation on CIFAR-10, the training accuracy, the testing accuracy and the highest attack advantage are $0.620$, $0.551$ and $0.003$, respectively. However, the testing accuracy of AlexNet trained without any defense on CIFAR-10 is $0.653$, which is $0.102$ higher than the testing accuracy of a model trained with knowledge distillation.

\subsection{Varying size of training set}  
\label{sec:vary_num}
\begin{figure*}
    \centering
    \includegraphics[width=16cm,height=7.5cm]{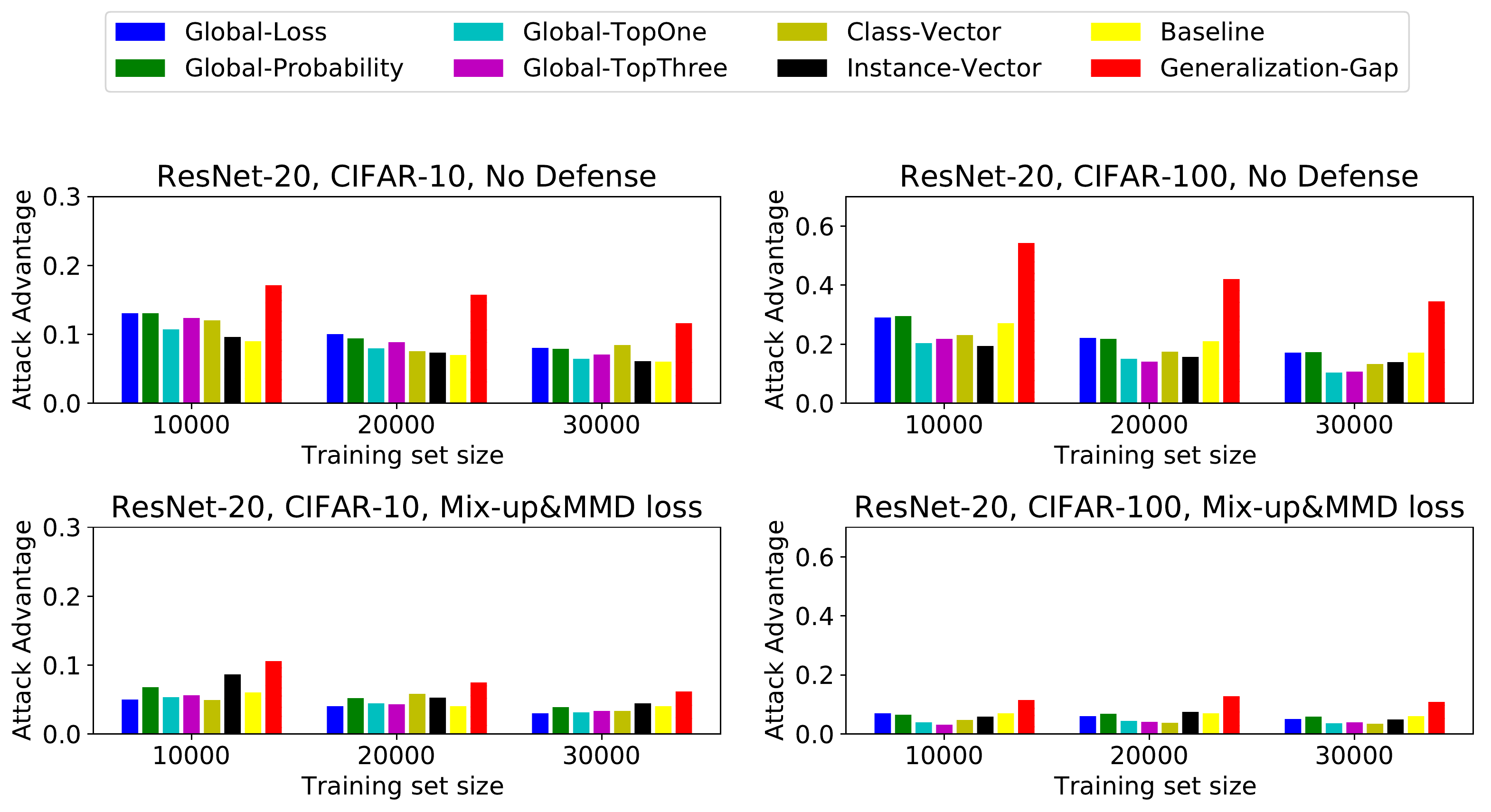}
    \caption{Attack advantage vs size of training set. ResNet-20 on CIFAR-10 and CIFAR-100 dataset, with and without MMD+Mix-up defense.}
    \label{num_attack}
\end{figure*}

Intuitively, a classifier's vulnerability to MI attacks is related to the size of the training set.  Here we explore the effect of different size of training set by using three different sizes: $\{10000,20000,30000\}$ on CIFAR-10 and CIFAR-100 datasets, with ResNet-20. 
The two plots in the first row of Figure~\ref{num_attack} shows the results.  We have three sets of attack advantage bars for all three different training set sizes. We see that the attack advantage values drop when the size of training set is increasing. However, the drop is relatively small. When the size of training set is increased from 10000 to 30000, the maximum attack advantage drops from from $0.130$ to $0.084$ and $0.317$ to $0.194$ on CIFAR-10 and CIFAR-100 dataset, respectively. The two plots in the second row in Figure~\ref{num_attack} shows the effect of different number of training instances under the defenses. We can see that the attack advantage drops gradually when more training instances are used; however, the reduction effect is less than deploying the defenses. One needs to notice that adding more training instances can be combined with deploying defenses.

\subsection{Ablation study on our proposed defense}

\begin{figure*}
     \centering
     \includegraphics[width=18cm,height=11cm]{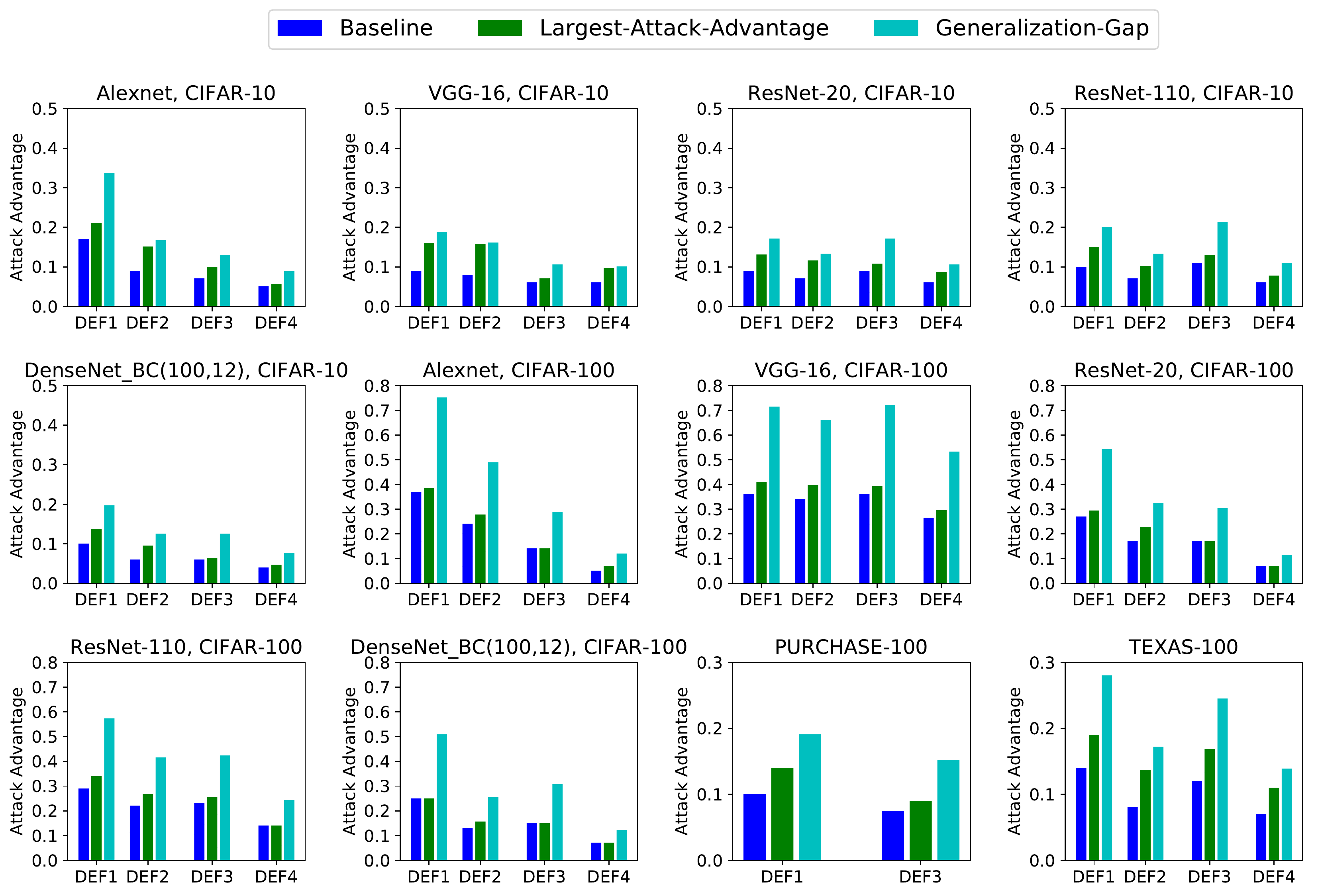}
     \caption{Attack advantage of different attacks for different classifiers on different datasets with different level of defenses. DEF1 means no defense, DEF2 means mix-up defense, DEF3 means MMD loss defense and DEF4 means MMD loss \&mix-up defense. }
     \label{fig:defense}
 \end{figure*}

In Figure \ref{fig:defense}, we compare the attack advantage of baseline attack, the highest attack advantage of existing attacks and the generalization gap under different defenses. One main observation is that using MMD loss or mix-up alone each offers some reduction in generalization gap; and combining mix-up with MMD loss offers better defense.  However, the generalization gap are not fully closed by the defense. We summarize the numbers presented in Figure \ref{fig:defense} in Table \ref{defense_avg_cifar10} and Table \ref{defense_avg_cifar100} for an average case comparison.

\subsection{The effect of our defense on prediction distributions}
\label{sec:distribution_change}

\begin{figure*}
        \centering
        \includegraphics[width=18cm,height=7cm]{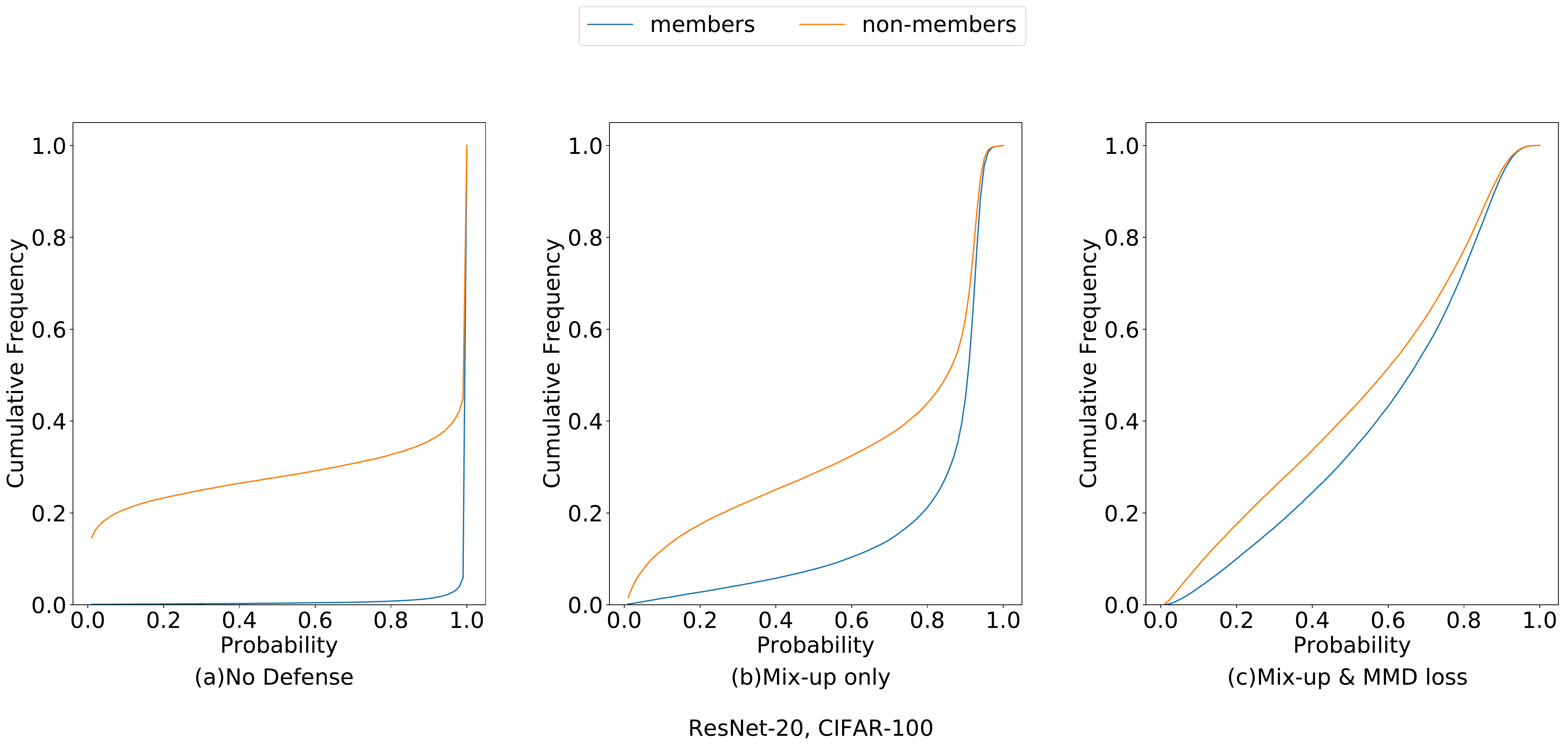}
        \caption{The distribution of the probability of correct label for members and non-members with different levels of defenses. ResNet-20, CIFAR-100.}
        \label{distribution_change}
\end{figure*}

To illustrate why the MMD+Mixup defense could work, we show the effect of our defenses on distributions of the probability of correct labels for members and non-members in Figure~\ref{distribution_change}. The neural network used here is ResNet-20 and the datasets is CIFAR-100. As shown in Figure~\ref{distribution_change} (a), when no defense is deployed, the two distributions are highly different from each other. The probability of correct label for members are mostly centered near 1. However, for non-members, a noticeable fraction of this group are clustered near 0, which means the predictions for this fraction are wrong. When mix-up defense is added, the two distributions become similar but the difference is still observable. After adding MMD loss defense, we see that the two distributions become almost the same. The changes on distributions explain why our defenses work. However, one still needs to notice that the generalization gap is not fully eliminated and the generalization gap is still larger than $0.1$ when all defenses are deployed. This means the attacker can still gain some attack advantage on the target model. 

   \begin{table*}
        \footnotesize
        \centering
        \begin{tabular}{|c|c|c|c|c|c|c|c|c|}
            \hline
             dataset& Model & learning rate & epochs & schedule & batch size & train acc.(\%) & test acc.(\%) & MMD loss weight \\
            \hline
              CIFAR-10 & AlexNet & 0.01 & 160 & [80,120] & 128 & 99.22 & 76.26 & 3\\
            \hline
             CIFAR-10 & VGG-16  & 0.01 & 160 & [80,120] & 128 & 99.31 & 91.77 & 1\\
            \hline
             CIFAR-10 & ResNet-20 & 0.1 & 160 & [80,120] & 128 & 99.39 & 90.62 & 0.5 \\
            \hline
             CIFAR-10 & ResNet-110 & 0.1 & 160 & [80,120] & 128 & 99.91 & 93.28 & 0.5 \\
            \hline
             CIFAR-10 & DenseNet\_BC(100,12) & 0.1 & 300 & [150,225] & 64 & 99.89 & 95.41 & 4\\ 
             
            \hline
             CIFAR-100 & AlexNet & 0.01 & 160 & [80,120] & 128 & 99.22 & 42.74 & 4\\
            \hline
             CIFAR-100 & VGG-16  & 0.01 & 160 & [80,120] & 128 & 99.81 & 69.91 & 0.3 \\
            \hline
             CIFAR-100 & ResNet-20 & 0.1 & 160 & [80,120] & 128 & 99.19 & 67.40 & 4\\
            \hline
             CIFAR-100 & ResNet-110  & 0.1 & 160 & [80,120] & 128 & 99.71 & 71.82 & 3.5\\
            \hline
             CIFAR-100 & DenseNet\_BC(100,12) & 0.1 & 300 & [150,225] & 64 & 99.79 & 76.13 & 2.5\\ 
            \hline
             PURCHASE-100 & PURCHASE-100 &0.001 & 100 & none & 100 & 98.12 & 79.04 & 0.1\\
            \hline
             TEXAS-100 & TEXAS-100 & 0.001 & 100 & none & 100 & 81.52 & 53.48 & 1\\
             \hline
             MNIST & MNIST & 0.01 & 100 & none & 128 & 99.67 & 99.03 & N/A\\
            \hline
        \end{tabular}
        \caption{Training recipe and model performance for different models. Learning rate is adjusted to 0.1x when current epoch is in schedule. Notice that ResNet-18 and DenseNet-121 are designed for ImageNet dataset, therefore these three models start with a convolutional layer with kernel size 7*7, which leads to slightly worse performance than its sibling CIFAR version: ResNet-20 and DenseNet\_BC(100,12).}
        \label{training_recipe}
    \end{table*}

\begin{table*}
    \normalsize
    \centering
    \begin{tabular}{lrrrr}
        & No Defense & Mix-up Alone (ablation) & MMD Alone (ablation) & MMD+Mix-up \\
        \toprule
       \cmidrule(r){1-5}
        Training accuracy \ & 0.990 & 0.944 & 0.909 & {\bf 0.883} \\
        Testing accuracy \ & 0.775 & {\bf 0.799} & 0.726 & 0.775 \\
      \cmidrule(lr){1-5}
        \textbf{Generalization\ gap} & 0.215 & 0.145 & 0.183 & {\bf 0.108} \\
        \textbf{Largest attack advantage} & 0.158 & 0.124 & 0.104 & {\bf 0.088} \\
        \textbf{Baseline attack advantage} & 0.112 & 0.074 & 0.096 & {\bf 0.061}  \\
      \cmidrule(lr){1-5}
        Class-Vector attack advantage  & 0.138 & 0.096 & 0.096 & {\bf 0.052} \\
        Global-Loss attack advantage  & 0.158 & 0.072 & 0.104 & {\bf 0.056} \\
        Global-Prob. attack advantage & 0.147 & 0.097 & 0.103 & {\bf 0.062} \\
        Global-TopOne attack advantage  & 0.113 & 0.082 & 0.062 & {\bf 0.047}  \\
        Global-TopThree attack advantage & 0.131 & 0.083 & 0.065 & {\bf 0.049} \\
        Instance-Vector attack advantage & 0.111 & 0.124 &  0.087 & {\bf 0.088} \\
        \bottomrule
    \end{tabular}
    \caption{(Ablation study) Average Training/Testing accuracy and average attack advantage, CIFAR-10 dataset.}
    \label{defense_avg_cifar10} 
    
    \vspace{2em}
    
    \centering
    \begin{tabular}{lrrrr}
         & No Defense & Mix-up Alone (ablation) & MMD Alone (ablation)& MMD+Mix-up   \\
                \toprule
 
        Training accuracy & 0.985 & 0.849 & 0.703 & {\bf 0.592} \\
        Testing accuracy & 0.366 & {\bf 0.421} & 0.284 & 0.365 \\
        \cmidrule(lr){1-5}
        \textbf{Generalization gap} & 0.619 & 0.429 & 0.419 & {\bf 0.231}  \\
        \textbf{Largest attack advantage} & 0.318 & 0.265 & 0.214  & {\bf0.133}\\
        
        \textbf{Baseline attack advantage} & 0.308 & 0.221 & 0.216 & {\bf 0.118} \\
        \cmidrule(lr){1-5}
        Class-Vector attack advantage& 0.299 & 0.157 & 0.144 & {\bf 0.088} \\
        Global-Loss attack advantage& 0.315 & 0.193 & 0.214 & {\bf 0.112} \\
        Global-Prob. attack advantage & 0.318 & 0.218 & 0.212 & {\bf 0.117} \\
        Global-TopOne attack advantage & 0.231 & 0.173 & 0.095 & {\bf 0.063}  \\
        Global-TopThree attack advantage & 0.251 & 0.174 & 0.103 & {\bf 0.068}\\
        Instance-Vector attack advantage & 0.211 & 0.265 & 0.130 & {\bf 0.133}  \\
        \bottomrule
    \end{tabular}
    \caption{(Ablation study) Average Training/Testing accuracy and average attack advantage, CIFAR-100 dataset.}
    \label{defense_avg_cifar100}
\end{table*}

